\documentclass[aps,prl,twocolumn,superscriptaddress]{revtex4-1}

\usepackage{mhchem,graphicx,longtable,xfrac}
\usepackage[colorlinks=true,citecolor=blue,linkcolor=blue,breaklinks=true]{hyperref}
\usepackage{amssymb}
\usepackage{graphicx}
\usepackage{amsmath}

\usepackage{times}
\usepackage{color}
\usepackage{bm}

\begin{document}

\newcommand {\beq} {\begin{equation}}
\newcommand {\eeq} {\end{equation}}
\newcommand {\bqa} {\begin{eqnarray}}
\newcommand {\eqa} {\end{eqnarray}}
\newcommand {\ba} {\ensuremath{b^\dagger}}
\newcommand {\Ma} {\ensuremath{M^\dagger}}
\newcommand {\psia} {\ensuremath{\psi^\dagger}}
\newcommand {\psita} {\ensuremath{\tilde{\psi}^\dagger}}
\newcommand{\lp} {\ensuremath{{\lambda '}}}
\newcommand{\A} {\ensuremath{{\bf A}}}
\newcommand{\Q} {\ensuremath{{\bf Q}}}
\newcommand{\kk} {\ensuremath{{\bf k}}}
\newcommand{\qq} {\ensuremath{{\bf q}}}
\newcommand{\kp} {\ensuremath{{\bf k'}}}
\newcommand{\rr} {\ensuremath{{\bf r}}}
\newcommand{\rp} {\ensuremath{{\bf r'}}}
\newcommand {\ep} {\ensuremath{\epsilon}}
\newcommand{\nbr} {\ensuremath{\langle ij \rangle}}
\newcommand {\no} {\nonumber}
\newcommand{\up} {\ensuremath{\uparrow}}
\newcommand{\dn} {\ensuremath{\downarrow}}
\newcommand{\rcol} {\textcolor{red}}



\title{Frustrated Heisenberg $J_1$\---$J_2$ model within the stretched diamond lattice of LiYbO$_2$}

\author{Mitchell M. Bordelon}
\affiliation{Materials Department, University of California, Santa Barbara, California 93106, USA}
\altaffiliation{Contributed equally to this work}

\author{Chunxiao Liu}
\affiliation{Department of Physics, University of California, Santa Barbara, California 93106, USA}
\altaffiliation{Contributed equally to this work}

\author{Lorenzo Posthuma}
\affiliation{Materials Department, University of California, Santa Barbara, California 93106, USA}

\author{Eric Kenney}
\affiliation{Department of Physics, Boston College, Chestnut Hill, Massachusetts 02467, USA}

\author{M. J. Graf}
\affiliation{Department of Physics, Boston College, Chestnut Hill, Massachusetts 02467, USA}

\author{N. P. Butch}
\affiliation{NIST Center for Neutron Research, National Institute of Standards and Technology, Gaithersburg, Maryland 20899, USA}

\author{Arnab Banerjee}
\affiliation{Neutron Scattering Division, Oak Ridge National Laboratory, Oak Ridge, TN 37831, USA}
\affiliation{Purdue Quantum Science and Engineering Institute, Purdue University, West Lafayette IN - 47906}

\author{Stuart Calder}
\affiliation{Neutron Scattering Division, Oak Ridge National Laboratory, Oak Ridge, TN 37831, USA}

\author{Leon Balents}
\affiliation{Kavli Institute for Theoretical Physics, University of California, Santa Barbara, Santa Barbara, California 93106, USA}

\author{Stephen D. Wilson}
\email[]{stephendwilson@ucsb.edu}
\affiliation{Materials Department, University of California, Santa Barbara, California 93106, USA}


\date{\today}

\begin{abstract}

  We investigate the magnetic properties of LiYbO$_2$, containing a
  three-dimensionally frustrated, diamond-like lattice via neutron
  scattering, magnetization, and heat capacity measurements. The
  stretched diamond network of Yb$^{3+}$ ions in LiYbO$_2$ enters a
  long-range incommensurate, helical state with an ordering wave
  vector ${\bf{k}} = (0.384, \pm 0.384, 0)$ that ``locks-in'' to a
  commensurate ${\bf{k}} = (1/3, \pm 1/3, 0)$ phase under the
  application of a magnetic field. The spiral magnetic ground state of
  LiYbO$_2$ can be understood in the framework of a Heisenberg
  $J_1$\---$J_2$ Hamiltonian on a stretched diamond lattice, where the
  propagation vector of the spiral is uniquely determined by the ratio
  of $J_2/|J_1|$.  The pure Heisenberg model, however, fails to
  account for the relative phasing between the Yb moments on the two
  sites of the bipartite lattice, and this detail as well as the
  presence of an intermediate, partially disordered, magnetic state
  below 1 K suggests interactions
  beyond the classical Heisenberg description of this material.

\end{abstract}

\maketitle

\section{I. Introduction}

In the field of three-dimensionally frustrated magnets, the predominant research focus has centered on the magnetic diamond and pyrochlore lattices \cite{bergman-balents, lee-balents, buessen-trebst, gchen, bernier2008quantum, savary2011impurity, bramwell1998frustration, harris1996frustration, harris1998liquid, moessner1998properties, canals1998pyrochlore, ramirez1999zero, bramwellspin, bramwell2001spinice, gardner2010magnetic, ross2011quantum}. Both of these frameworks appear within the family of transition-metal spinels of the form $AB_2X_4$ ($A,B$ = transition metal or metalloid, $X$ = chalcogenide), where the diamond and pyrochlore lattices appear on the $A$- and $B$-site sublattices, respectively. Strong magnetic frustration within each of these sublattice types is known to suppress typical Ne\'el order and instead favor the manifestation of unconventional ground states, including classical spin liquids \cite{moessner1998properties, canals1998pyrochlore}, (quantum) spin ices \cite{ramirez1999zero, bramwellspin, bramwell2001spinice, ross2011quantum}, and (quantum) spiral spin liquids \cite{bergman-balents, lee-balents, buessen-trebst}. 

Quantum fluctuations that manifest in the small spin limit on these lattices further suppress magnetic order and can formulate the basis for highly entangled ground states \cite{Lee, balents4, balents5, witczak6, zhou7, cava-broholm}. At this limit, the magnetic diamond lattice has been less thoroughly studied in comparison to the magnetic pyrochlore lattice, as the magnetic pyrochlore lattice also manifests in a large, well-studied family of rare-earth $Ln_2M_2O_7$ ($Ln$ = lanthanide, $M$ = metal or metalloid) compounds \cite{bernier2008quantum, savary2011impurity, bramwell1998frustration, harris1996frustration, harris1998liquid, moessner1998properties, canals1998pyrochlore, ramirez1999zero, bramwellspin, bramwell2001spinice, gardner2010magnetic, ross2011quantum}.  Furthermore, while introducing model $J_{eff}=1/2$ lanthanide moments within frustrated magnetic motifs has shown promise in realizing intrinsically quantum disordered states (e.g. Yb$_2$Ti$_2$O$_7$ pyrochlore \cite{ross2011quantum, gaudet-gaulin} and triangular lattice NaYbO$_2$ \cite{bordelon2019field, bordelon2020, ding-tsirlin, ranjith-baenitz}), isolating materials that comparably incorporate model $f$-electron moments within a diamond lattice framework is a challenge.

\begin{figure*}[t]
	\includegraphics[width=\textwidth*10/10]{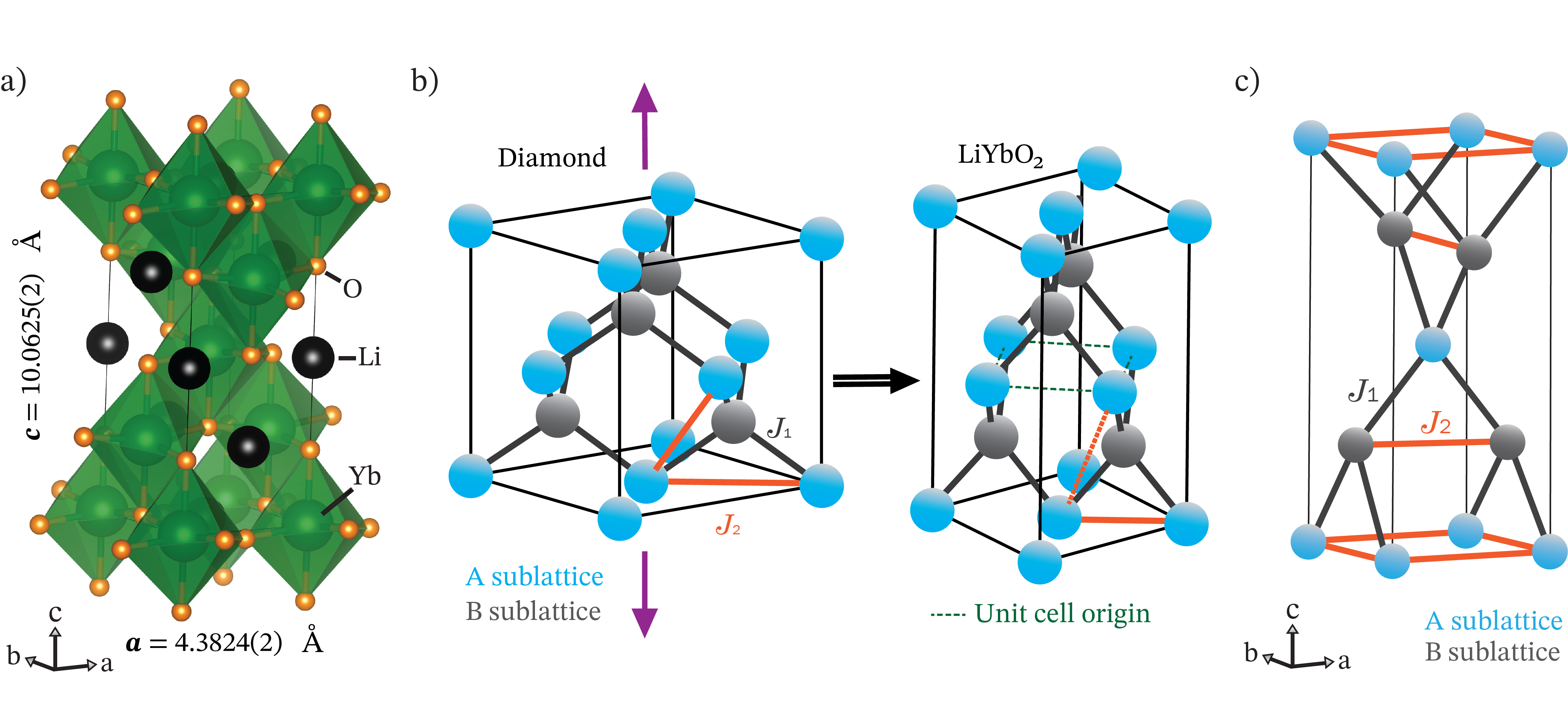}
	\caption{a) Crystal structure of LiYbO$_2$ with YbO$_6$ octahedra shaded in green and black spheres noting the positions of Li ions.  b) The frustrated $J_1 – J_2$ model on the diamond lattice consists of two interpenetrating face centered cubic (FCC) sublattices, A and B, with a $J_1$ (black) magnetic interaction connecting the two sublattices and a $J_2$ (orange) spanning interactions within an FCC sublattice. When this structure is stretched along one of the cubic axes, the $I4_1/amd$ lattice of LiYbO$_2$ is reproduced where the dashed green line represents the unit cell origin of LiYbO$_2$ shown in panel c). In LiYbO$_2$, the stretched bond (5.909 \AA, dashed orange) is 1.527 \AA \ longer than the in-plane $J_2$ (4.382 \AA, solid orange). In the present model for LiYbO$_2$, the stretched bond is assumed negligible in strength relative to the shorter $J_2$. c)  NN ($J_1$) and NNN ($J_2$) exchange pathways between Yb-ions in LiYbO$_2$ with Yb ions in the $A$ and $B$ sublattices shaded differently for clarity.}
	\label{fig:fig1}
\end{figure*}

Frustration within the diamond lattice is best envisioned by dividing the lattice into two interpenetrating face centered cubic (FCC) lattices with two exchange interactions, $J_1$ and $J_2$, where in the Heisenberg limit (Figure \ref{fig:fig1}) \cite{bergman-balents, lee-balents, buessen-trebst}.
\begin{equation}
	\label{eq:J1J2}
	H = J_1\sum_{<i,j>} {\bf{S_i}}\cdot{\bf{S_j}} + J_2\sum_{<<i,j>>} {\bf{S_i}}\cdot{\bf{S_j}}
\end{equation}
In the two limits where either $J_1$ or $J_2$ is zero, this bipartite system is unfrustrated with a conventional Ne\'el ordered ground state. However, when $J_2 > 0$ and $|J_1| > 0$, ordering becomes frustrated. When $J_2 /|J_1| \ge 1/8$, the classical interpretation of this model develops a degenerate ground state manifold of coplanar spin spirals \cite{bergman-balents, lee-balents, buessen-trebst}. Each of these spirals can be described by a unique momentum vector, and together the degenerate momentum vectors formulate a spin spiral surface in reciprocal space \cite{bergman-balents, lee-balents, buessen-trebst}. The degeneracy of these spin spirals can be lifted entropically via an order-by-disorder mechanism that selects a unique spin spiral state \cite{bergman-balents, lee-balents, buessen-trebst}, but in the presence of strong quantum fluctuations ($S \le 1$), long-range order is quenched and a spiral spin liquid ground state manifests that fluctuates about the spiral surface \cite{buessen-trebst}.

Identifying materials exhibiting (quantum) spiral spin liquid states derived from this $J_1$\---$J_2$ model remains an outstanding goal.  Transition-metal-based $AB_2X_4$ spinels have been primarily investigated as potential hosts; however two vexing problems typically occur: (1) non-negligible further neighbor interactions beyond the $J_1$\---$J_2$ limit arise and lift the degeneracy and (2) weak tetragonal distortions from the ideal $Fm\bar{3}m$ spinel structure appear. For example, detailed investigations of the spinels MgCr$_2$O$_4$ \cite{tomiyasu-yamada, bai-mourigal}, MnSc$_2$S$_4$ \cite{gao-ruegg, iqbal-reuther, krimmel-loidl}, NiRh$_2$O$_4$ \cite{buessen-trebst, chamorro-mcqueen}, and CoRh$_2$O$_4$ \cite{flynn-mourigal} have all required expanding the model Hamiltonian to include up to third-neighbor interactions, originating from the large spatial extent of $d$-orbitals, to describe the generation of their helical magnetic ground states. Within some materials like NiRh$_2$O$_4$ \cite{buessen-trebst, chamorro-mcqueen}, single ion anisotropies must also be incorporated to digest the experimental results. Complexities with extended interactions beyond the $J_1$\---$J_2$ limit may also compound with inequivalent exchange pathways that form as the cubic $Fm\bar{3}m$ spinel structure undergoes a distortion to a tetragonal $I4_1/amd$ or $I\bar{4}2d$ space group prior to magnetic ordering (e.g. NiRh$_2$O$_4$ \cite{buessen-trebst, chamorro-mcqueen} and CoRh$_2$O$_4$ \cite{flynn-mourigal}). 

The tetragonal distortion in spinels can be viewed as a compression of the diamond lattice along one of its cubic axes (opposite to that illustrated in Figure \ref{fig:fig1}), and it splits the nominal $J_2$ of the ideal diamond lattice structure into two different pathways. This disrupts the reciprocal space spiral surface generated in the $J_1$\---$J_2$ model's cubic limit. Despite these complications common to $A$-site transition metal spinels, the predictions born from the model Hamiltonian show substantial promise as materials such as MnSc$_2$S$_4$ \cite{gao-ruegg, iqbal-reuther, krimmel-loidl}, CoAl$_2$O$_4$ \cite{zaharko-loidl, roy-furukawa, macdougall-nagler}, and NiRh$_2$O$_4$ \cite{buessen-trebst, chamorro-mcqueen} are nevertheless either close to or partially manifest degenerate spiral spin states.  Identifying other crystal structures that realize comparable physics but with more localized $f$-electron moments is an appealing path forward.





Here we present an investigation of an alternative, frustrated diamond lattice framework in the material LiYbO$_2$. This material can be viewed as containing a stretched diamond lattice of Yb$^{3+}$ moments (Figure \ref{fig:fig1}), and it falls within a broader family of $ALnX_2$ ($A$ = alkali, $Ln$ = lanthanide, $X$ = chalcogenide) materials where the lattice structure is dictated by the ratio of lanthanide ion radius to alkali plus chalcogenide radii (Figure \ref{fig:fig2}).  Our results show that LiYbO$_2$ realizes the expected ground state derived from a $J_1$\---$J_2$ Heisenberg model on a tetragonally-elongated diamond lattice and that  $J_{eff} = 1/2$ Yb$^{3+}$ ions in related materials may act as the basis for applying the Heisenberg $J_1$\---$J_2$ model to $Ln$-ion diamond-like materials.  Notably, however, variance between the observed and predicted phasing of Yb moments on the bipartite lattice as well as the emergence of an intermediate, partially disordered state suggests the presence of interactions/fluctuation effects not captured in the classical $J_1$\---$J_2$ Heisenberg framework.

\section{II. Methods}

\subsection{Sample preparation}
Polycrystalline LiYbO$_2$ was prepared from Yb$_2$O$_3$ (99.99\%, Alfa Aesar) and Li$_2$CO$_3$ (99.997\%, Alfa Aesar) via a solid-state reaction in a 1:1.10 molar ratio. This off-stoichiometric ratio was used to compensate for the partial loss of Li$_2$CO$_3$ during the open crucible reaction.  The constituent precursors were ground together, heated to 1000 $^{\circ}$C for three days in air, reground, and then reheated to 1000 $^{\circ}$C for 24 hrs. Samples were kept in a dry, inert environment to prevent moisture absorption. Measurements were conducted with minimal atmospheric exposure to maintain sample integrity.  Sample composition was verified via x-ray diffraction measurements on a Panalytical Empyrean powder diffractometer with Cu-K$\alpha$ radiation, and data were analyzed using the Rietveld method in the Fullprof software suite \cite{rodriguezfullprof}.
  
\subsection{Magnetic susceptibility}
The bulk magnetization and static spin susceptibility of LiYbO$_2$ were measured using three different instruments. Low-field d.c. magnetization data from 2 to 300 K were collected on a Quantum Design Magnetic Properties Measurement System (MPMS3) with a 7 T magnet, and isothermal d.c. magnetization data between 2 to 300 K were collected on a Quantum Design Physical Properties Measurement System (PPMS) equipped with a vibrating sample magnetometer insert and a 14 T magnet. Low-temperature a.c. susceptibility data between 2 K and 330 mK were collected on an a.c. susceptometer at 711.4 Hz with a 0.1 Oe (7.96 A m$^{-1}$) drive field) in a $^{3}$He insert. The background generated by the sample holder in this low temperature a.c. measurement is subtracted from the data presented. 

\subsection{Heat capacity}
Specific heat measurements were collected between 100 mK and 300 K on sintered samples of LiYbO$_2$ in external magnetic fields of 0, 3, 5, and 9 T. Specific heat data between 2 to 300 K were collected on a Quantum Design PPMS with the standard heat capacity module, while specific heat data below 2 K was obtained with a dilution refrigerator insert. The lattice contribution to the specific heat of LiYbO$_2$ was modeled with a Debye function using two Debye temperatures of $\Theta_{D1} = 230.5$ K and $\Theta_{D2} = 615.3$ K. The magnetic specific heat was then obtained by subtracting out the modeled lattice contribution from the data, and $C_{mag}/T$ was integrated from 100 mK to 40 K to determine magnetic entropy of LiYbO$_2$ at 0, 3, 5, and 9 T.
  
\subsection{Neutron diffraction} 
Neutron powder diffraction data were collected on the HB-2A diffractometer at the High Flux Isotope Reactor (HFIR) in Oak Ridge National Laboratory. The sample was placed inside a cryostat with a $^{3}$He insert and a 5 T vertical field magnet, and data were collected between 270 mK and 1.5 K. Sintered pellets of LiYbO$_2$ were loaded into Cu canisters, and incident neutrons of wavelength $\lambda=2.41$ \AA ~were selected using a Ge(113) monochromator. Rietveld refinement of diffraction patterns was conducted using the FullProf software suite \cite{rodriguezfullprof}, and magnetic symmetry analysis was performed with the program $SARAh$ \cite{wills2000new}. The structural parameters were determined using data collected at 1.5 K and then fixed for the analysis of the temperature-subtracted data used for magnetic refinements. 

Inelastic neutron scattering (INS) data were collected on two instruments. High-energy inelastic data were obtained on the wide Angular-Range Chopper Spectrometer (ARCS) at the Spallation Neutron Source in Oak Ridge National Laboratory. Two incident neutron energies of $E_i$ = 150 meV (Fermi 2, Fermi frequency 600 Hz) and 300 meV (Fermi 1, Fermi frequency 600 Hz) were used, and data were collected at 5 K and 300 K \cite{lin2019energy}. Background contributions from the aluminum sample can were subtracted out by measuring an empty cannister under the same conditions. Crystalline electric field (CEF) analysis was conducted by integrating energy cuts ($E$-cuts) of the 300 meV data between $|Q| = [4, 6]$ \AA$^{-1}$. Integrated $E$-cuts of the 150 meV data between $|Q| = [2, 3]$ \AA$^{-1}$ are shown in the Supplementary Materials \cite{supplementalmaterial}. Peaks were fit with a Gaussian function that approximates the beam shape of the instrument. Low-energy inelastic scattering data were collected on the Disc Chopper Spectrometer (DCS) instrument at the NIST Center for Neutron Research (NCNR), National Institute of Standards and Technology (NIST). Neutrons of incident energy $E_i = 3.32$ meV in the medium-resolution setting were used, and the sample was loaded into a cryostat with a 10 T vertical field magnet and a dilution insert. 

\subsection{Crystalline electric field analysis} 
The crystalline electric field (CEF) of LiYbO$_2$ was fit following a procedure outlined in Bordelon et al. \cite{bordelon2020}, and a rough overview is reviewed here.

In LiYbO$_2$, magnetic Yb$^{3+}$ with total angular momentum $J = 7/2$ ($L = 3$, $S = 1/2$) is split into a series of four Kramers doublets in the local $D_{2d}$ CEF point group symmetry. Estimations of the splitting can be modeled with a point charge (PC) model of varying coordination shells in the crystal field interface of Mantid Plot \cite{Mantid}. Three coordination-shell variants with increasing distance from a central Yb ion are displayed as PC 1, PC 2, and PC 3 in Table \ref{tab:tab2}. The minimal Hamiltonian with CEF parameters $B_n^m$ and Steven's operators $\hat{O}_m^n$ \cite{StevensOperators} in $D_{2d}$ symmetry is written as follows:

\begin{multline}
	\label{eq:CEF}
	H_{CEF} = B_2^0 \hat{O}_2^0 + B_4^0 \hat{O}_4^0 + B_4^4 \hat{O}_4^4 + B_6^0 \hat{O}_6^0 + B_6^4 \hat{O}_6^4
\end{multline}

The diagonalized CEF Hamiltonian was used to calculate energy eigenvalues, relative transition intensities, a powder-averaged $g_{avg}$ factor, and corresponding wave functions. These values were compared with data obtained from integrated $E$-cuts of ARCS 300 meV data and bulk magnetic property measurements. The deviation was minimized with a combination of Mantid Plot \cite{Mantid}, SPECTRE \cite{spectre}, and numerical error minimization according to the procedure in Bordelon et al. \cite{bordelon2020, gaudet-gaulin} to approach a global minimum that represents the Yb CEF environment in LiYbO$_2$. 

\section{III. Experimental Results}

\begin{figure}[t]
	\includegraphics[scale=.55]{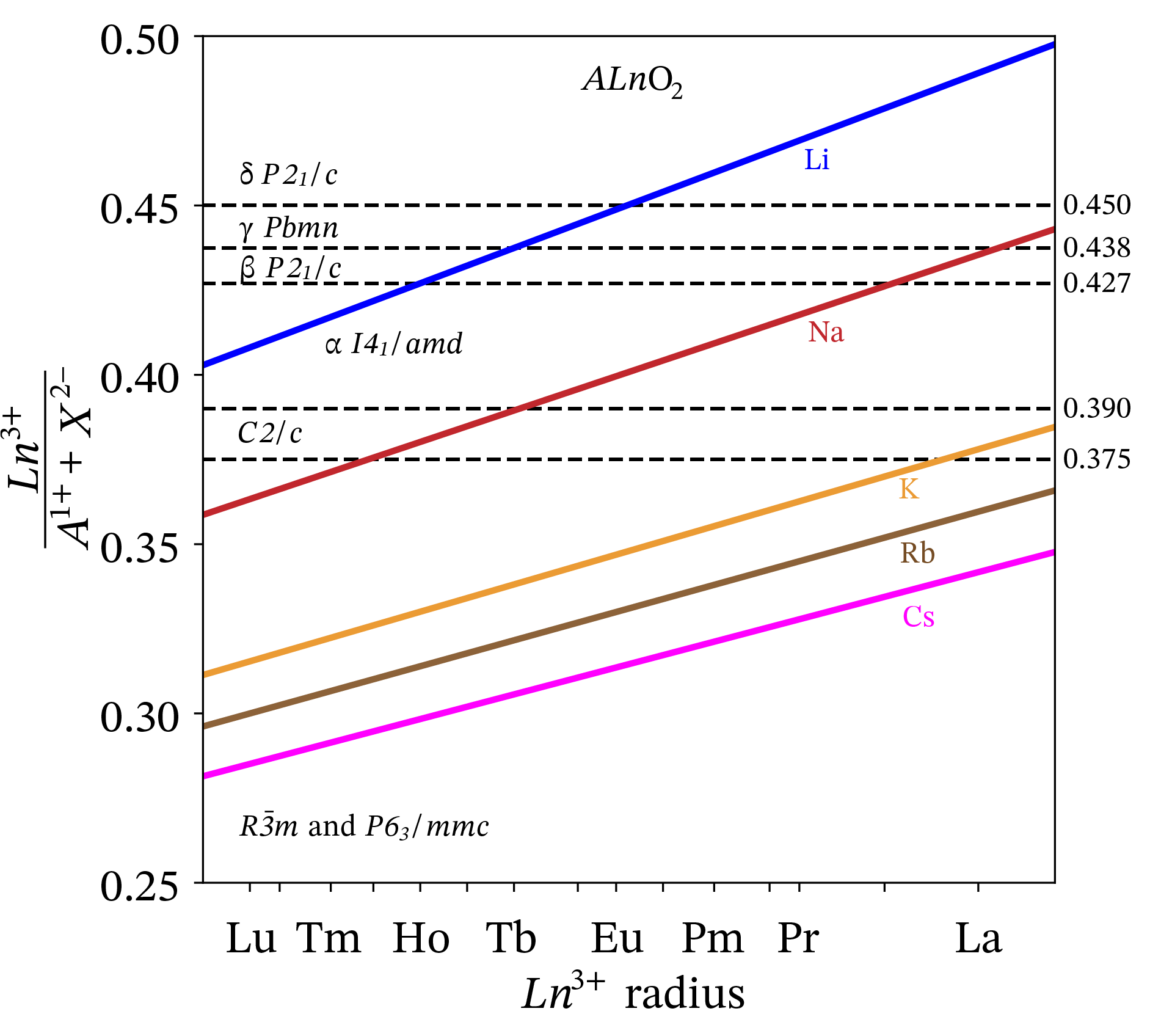}
	\caption{The series of $ALnX_2$ ($A$ = alkali, $Ln$ = lanthanide, $X$ = chalcogenide) compounds crystallize in several structures governed by the ratio of lanthanide radius divided by the sum of the alkali and chalcogenide radii derived from tabulated ionic radii and reported crystal structures \cite{shannon-prewitt, shannon-prewitt-revised, hashimoto2002structures, hashimoto2003magnetic, bronger1993synthesis, dong2008structure, cantwell2011crystal, liu-zhangAMX2, xing-sefat2}. Crossover between differing phases occurs at the dashed lines, and materials on these lines can crystallize in both neighboring space groups (e.g. NaErO$_2$ $R\bar{3}m$ and $C2/c$ \cite{hashimoto2003magnetic}).}
	\label{fig:fig2}
\end{figure}

\subsection{Radius ratio rule in $\bf{ALnX_2}$ materials}

The $I4_1/amd$ space group is one of the seven major space groups ($R\bar{3}m$ and $P6_3/mmc$, $C2/c$, $\alpha$-$I4_1/amd$, $\beta$-$P2_1/c$ $\gamma$-$Pbmn$, $\delta$-$P2_1/c$) that represent the $ALnX_2$ compounds as shown in Figure \ref{fig:fig2}. The structure types adopted by this family of compounds switch depending on the relative sizes between alkali and lanthanide radii. An empirical relationship between the radii of all three chemical constituents of the $ALnX_2$ family and the major space groups reported in this series is shown in Figure \ref{fig:fig2} by comparing reported structures in the literature \cite{hashimoto2002structures, hashimoto2003magnetic, bronger1993synthesis, dong2008structure, cantwell2011crystal, liu-zhangAMX2, xing-sefat2} to tabulated ionic radii \cite{shannon-prewitt, shannon-prewitt-revised}. The $\alpha, \beta, \gamma, \delta$ follow the nomenclature of  Hashimoto et al. \cite{hashimoto2002structures}, and the $R\bar{3}m$ space group is the $\alpha$-NaFeO$_2$ structure type. Plots of the radius ratio relationships for varying chalcogenides are also displayed in the Supplemental Material section \cite{supplementalmaterial}.

Compounds residing close to or on the dashed lines separating two space groups can crystallize in either space group depending on synthesis conditions or temperature. For example, NaErO$_2$ crystallizes in $R\bar{3}m$ and $C2/c$ \cite{hashimoto2003magnetic} structures at room temperature, and LiErO$_2$ goes through a structural phase transition from $\alpha$-$I4_1/amd$ at 300 K to $\beta$-$P2_1/c$ at 15 K \cite{hashimoto2002structures}. Two related crystal structures are possible in the $R\bar{3}m$ and $P6_3/mmc$ area of Figure \ref{fig:fig2}, and both of these space groups contain sheets of equilateral triangles comprised of lanthanide ions and vary only in the stacking sequence of the triangular sheets ($ABC$ for $R\bar{3}m$ and $AAA$ for $P6_3/mmc$). Previous reports also indicate that the $P6_3/mmc$ phase is favored with large Cs$^+$ ions \cite{xing-sefat2, bronger1993synthesis}.  We note here that this empirical radius-ratio rule excludes one of the known $ALnX_2$ phases: the chemically-disordered $Fm\bar{3}m$ NaCl phase that is primarily present at high temperatures when the alkali radius is close to that of the lanthanide radius \cite{liu-zhangAMX2, bronger1973ueber, verheijen1975flux, tromme, ohtani1987synthesis, fabry2014structure}. This chemically-disordered phase goes through a first order phase transition to the $R\bar{3}m$ phase in materials such as NaNdS$_2$ \cite{ohtani1987synthesis, fabry2014structure}.

\begin{table}[t]
	\caption{Rietveld refinement structural parameters at 1.5\,K from elastic neutron scattering measurements on LiYbO$_2$ on HB2A in the $I4_1/amd$ space group with origin setting two. Within error, all ions refined to full occupation and no quantifiable site mixing is present.}
	\begin{tabular}{cc|ccccc}
		\hline
		\multicolumn{2}{c|}{$T$}       & \multicolumn{5}{c}{1.5 K}     \\ \hline
		\multicolumn{2}{c|}{\ $\chi^{2}$}       & \multicolumn{5}{c}{3.421}     \\ 
		\multicolumn{2}{c|}{$\lambda$} & \multicolumn{5}{c}{2.41 \AA}   \\
		\multicolumn{2}{c|}{$a=b$}     & \multicolumn{5}{c}{4.3824(2) \AA}  \\
		\multicolumn{2}{c|}{$c$}       & \multicolumn{5}{c}{10.0625(2) \AA}  \\ \hline
		Atom           & Wyckoff          & x    & y    & z            & $B_{iso}$ (\AA$^{2}$)      & Occupancy  \\ \hline
		Yb             & 4a            & 0    & 0    & 0            & 0.28(9)   & 1.000(6) \\
		Li             & 4b            & 0    & 0    & 0.5          & 2.02(30)   & 1.00(5) \\
		O              & 8e            & 0    & 0    & 0.22546(7)   & 0.74(9)   & 1.00(3) \\ \hline	
	\end{tabular}
	\label{tab:tab2}
	
\end{table}

\begin{figure*}[t]
	\includegraphics[width=\textwidth*10/10]{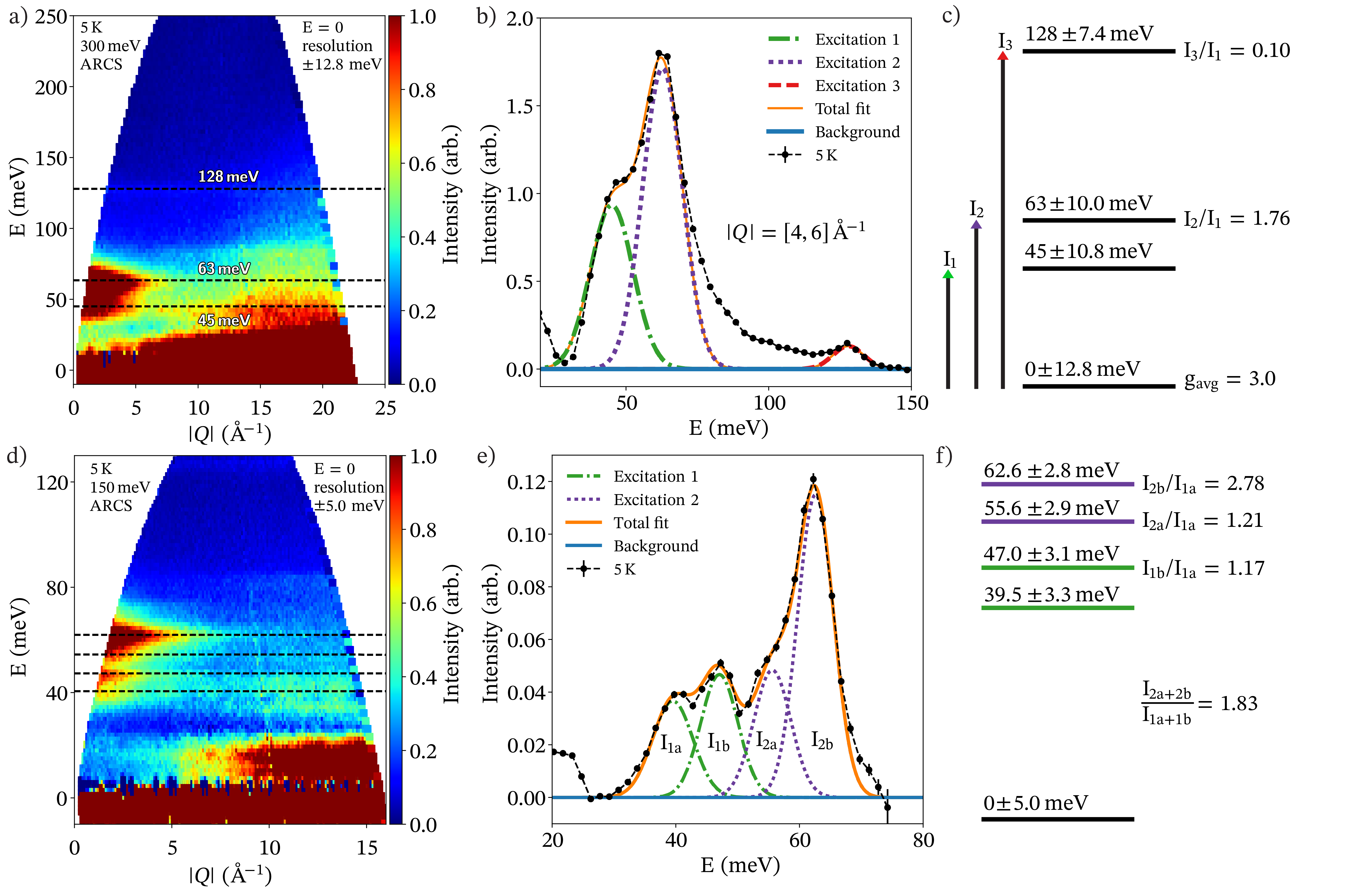}
	\caption{a) Inelastic neutron scattering (INS) spectrum $S(\bf{Q},\hbar \omega)$ collected at 5 K and $E_i = 300$ meV on the ARCS spectrometer with full width half max resolution in the elastic line of 12.8 meV. Three CEF levels indicated by dashed black lines were observed. b) $\bf{Q}$-integrated cut from panel a) overplotted with model lineshapes derived from the CEF fits in Table 1 and convolved with the instrumental resolution in MantidPlot \cite{Mantid}. c) The $D_{2d}$ $J = 7/2$ Yb$^{3+}$ ion generates four Kramers doublets centered at 45, 63, and 128 meV determined via CEF fits to the INS data. Errors shown correspond to instrumental resolution, and the well-separated ground state doublet has an average g-factor $g_{avg}=3.0$. The intensity ratios $I_2/I_1$ and $I_3/I_1$ were determined relative to the 45 meV CEF excitation. d) INS spectrum $S(\bf{Q},\hbar \omega)$ collected at 5 K and $E_i = 150$ meV from the ARCS spectrometer. The first two CEF excitations observed with $E_i = 300$ meV show splitting of $\approx7$ meV, highlighted by the dashed black lines. At $E_i = 300$ meV, this splitting is not resolvable due to the poorer energy resolution. e-f) Integrating the split CEF excitations shows that the ratio of integrated intensities of CEF mode 2 (I$_{2a}+I_{2b}$) to CEF mode 1 (I$_{1a}+I_{1b}$) correlates to the ratios of the integrated intensities in b). The splitting of CEF excitations may represent two distinct chemical environments within LiYbO$_2$ that are not resolvable within structural neutron powder diffraction experiments. }
	\label{fig:fig5}
\end{figure*}

\begin{table*}[]
	
	\caption{Point charge (PC) models and CEF fit for LiYbO$_2$ obtained by minimizing observed parameters from $E_i$ = 300 meV INS data and powder averaged $g_{avg}$ factor from isothermal magnetization. The three PC models of increasing size incorporate one (O$^{2-}$ ions), two (O$^{2-}$ and Li$^{+}$ ions) and three (O$^{2-}$, Li$^{+}$, and Yb$^{3+}$ ions) coordination shells surrounding a central Yb$^{3+}$ ion, respectively.}
	\begin{tabular*}{14.5cm}{l|llllll|l|llllll}
		\hline
		& $E_1$   & $E_2$    & $E_3$    & $\frac{I_2}{I_1}$  & $\frac{I_3}{I_1}$   & $g_{avg}$  & $\chi^2$ & $B_2^0$      &  $B_4^0$        & $B_6^0$         &  $B_4^4$      &  $B_6^4$     \\ \hline
		PC (2.5 \AA)  & 33.3 & 33.8 & 69.0 & 0.98 & 0.08 & 3.6 & 51.0 & -0.67210 & -0.031153 & 0.000064591 & -0.17420 & -0.0012000       \\
		PC (3.1 \AA)  & 30.9 & 86.2 & 87.5 & 0.10 & 0.06 & 3.7 & 27.8 & 2.1336 & -0.029755 & 0.000069724 & -0.19050 & -0.0012660
		\\
		PC (3.5 \AA)  & 108.6 & 149.9 & 156.6 & 0.07 & 0.08 & 4.5 & 232.0 & -4.2146 & -0.033288 & 0.000081398 & -0.18211 & -0.0014202       \\
		
		Fit  & 45.0 & 62.8 & 127.9 & 1.74 & 0.10 & 3.0 & 0.002 & 0.31777 & -0.072378 & 0.0010483 & -0.27051 & 0.0015364     \\ \hline 
		Observed   & 45.0 & 63.0 & 128.0 & 1.76 & 0.10 & 3.0  & & &  &  &  &  &  
	\end{tabular*}
	
	
	\hskip+0.1cm
	\begin{tabular*}{14.525cm}{ll}
		\hline
		Fit wave functions:     &  \\
		$| \omega_{0, \pm}\rangle =$  $0.901|\mp1/2\rangle +0.434|\pm7/2\rangle$ &\\
		$| \omega_{1, \pm}\rangle =$  $-0.434|\mp1/2\rangle +0.901|\pm7/2\rangle$ &\\
		$| \omega_{2, \pm}\rangle =$  $0.849|\pm3/2\rangle +0.529|\mp5/2\rangle$ &\\
		$| \omega_{3, \pm}\rangle =$  $-0.529|\mp3/2\rangle +0.849|\pm5/2\rangle$ &\\ \hline
	\end{tabular*}
	\label{tab:tab1}
	
\end{table*}

\subsection{Chemical structure}
Elastic neutron powder diffraction data collected from LiYbO$_2$ are shown in Figure \ref{fig:fig6}. The crystal structure was fit at 1.5 K to the $I4_1/amd$ structure previously reported \cite{hashimoto2002structures}, and this structural fit was used as the basis for analyzing the magnetic peaks observed below 1.5 K as a function of magnetic field. Details of the structural fit are presented in Table \ref{tab:tab2}. Within resolution of this experiment, all chemical sites are fully occupied without site-mixing, and no impurity phases are present. 

LiYbO$_2$ consists of $D_{2d}$ edge-sharing YbO$_6$ octahedra that are connected three-dimensionally within a bipartite magnetic lattice (Figure \ref{fig:fig1}). Each sublattice of trivalent Yb ions ($A$ or $B$ sublattice in Figure \ref{fig:fig1}) connects to the neighboring sublattice's layers with two bonds above and two bonds below with a nearest-neighbor Yb$_{A/B}$-Yb$_{B/A}$ distance of 3.336 \AA \ ($J_1$). This forms a stretched tetrahedron with a Yb ion at its center. The next-nearest-neighbor bond is within the same Yb sublattice where four bonds within the $ab$-plane are connected at 4.4382 \AA \ ($J_2$). Despite this nearest-neighbor and next-nearest-neighbor interaction appearing significantly different in length, superexchange is likely promoted along $J_2$ due to the more favorable Yb-O-Yb bond angle, making the longer next-nearest neighbor exchange comparably relevant to the nearest-neighbor $J_1$.  Exchange pathways through oxygen anions along $J_1$ and $J_2$ are nearly equivalent at 4.473 \AA \ and 4.410 \AA, respectively. Therefore, the two magnetic exchange interactions are likely similar in magnitude, and when $|J_1| > 0$ and $J_2 > 0$, this lattice is expected to be geometrically frustrated.

The Yb$^{3+}$ magnetic lattice can be visualized as an extreme limit of tetragonal elongation of the diamond lattice as shown in Figure \ref{fig:fig1}. The diamond lattice originally contains two magnetic interactions, $J_1$ and $J_2$, where $J_2$ interactions within any face of the diamond lattice are equivalent. Stretching the lattice in LiYbO$_2$ breaks the $J_2$ degeneracy, creating a $J_{2a}$ interaction along the elongated direction 5.9090 \AA \ and an in-plane $J_{2b}$ of 4.438 \AA. In the full chemical unit cell of LiYbO$_2$, the elongated $J_{2a}$ interaction necessitates two O$^{2-}$ ion superexchange links relative to the single O$^{2-}$ superexchange in the in-plane $J_{2b}$ and the $J_1$ interaction. As it is likely negligible in strength relative to the other two interactions, the elongated $J_{2a}$ interaction is therefore neglected in this paper and $J_{2b}$ is simply referred to as $J_2$. 

\subsection{Crystalline electric field excitations}
Inelastic neutron scattering (INS) data were collected at $T$ = 5 K and $E_i$ = 300 meV to map the intramultiplet CEF excitations in LiYbO$_2$.   Figures \ref{fig:fig5}a-c) show three CEF excitations that are centered around 45, 63, and 128 meV. A cut through $S(\bf{Q},\hbar \omega)$ shows the energy-widths of the transitions in Fig. 3b) are limited by the instrumental resolution at $E_i$ = 300 meV. As expected, the lowest-energy CEF transition is high enough to render the ground state Kramers doublet a well-separated $J_{eff} = 1/2$ state at low temperatures. An analysis of the CEF splitting of the $J = 7/2$ Yb$^{3+}$ manifold is detailed in Figure \ref{fig:fig5} and Table \ref{tab:tab1}. With the extracted parameters from the $S(\bf{Q},\hbar \omega)$ cut, the best level scheme fit to the data is shown in Table \ref{tab:tab1}. The calculated CEF $g_{avg}$ is split into two anisotropic components of $g_{//} = 0.58$ and $g_{\perp} = 3.71$, where $g_{avg} = \sqrt (1/3( g_{//}^2 + 2g_{\perp}^2))$. The fit diverges from point charge models of varying coordination size presented in Table \ref{tab:tab1} and is closest in sign to the parameters $B_m^n$ generated from a point charge model incorporating two ionic shells (3.1 \AA \ with O$^{2-}$ and Li$^{+}$ ions). 

The first two CEF excitations were further analyzed with lower $E_i $ = 150 meV INS data presented in Figure \ref{fig:fig5}d-f). Within this higher resolution window, the lower two CEF excitations at 45 meV and 63 meV show new features, and the two CEF excitations are asymmetrically split into peaks centered at 39.5 meV + 47.0 meV (excitation 1) and 55.6 meV + 62.6 meV (excitation 2). At $E_i $ = 300 meV, this splitting is below the instrumental resolution and is not readily apparent. The relative integrated intensities of the split modes in excitation 1 and excitation 2 at $E_i $ = 150 meV however agree with the ratios of the single/convolved modes observed in the $E_i $ = 300 meV. The most likely explanation for the observed splitting at $E_i $ = 150 meV is the presence of two distinct chemical environments surrounding Yb ions that are outside of the resolution of the current neutron powder diffraction measurements. 

LiYbO$_2$ indeed contains two sublattices of Yb ions ($A$ and $B$ in \ref{fig:fig1}c), and, in the ideal $I4_1/amd$ structure, Yb ions within each sublattice reside in chemically-equivalent environments. Since the CEF fit is closest to a point charge model including both nearest O$^{2-}$ and Li$^{+}$ ions, the observed splitting could arise from these non-magnetic ions residing slightly off of their ideal Wyckoff positions. A similar chemical feature has been observed in tetragonally-distorted spinels, such as CuRh$_2$O$_4$ \cite{flynn-mourigal} where Cu ions are displaced off of their ideal Wyckoff site. While such a feature is outside of the resolution of the average structural refinement for LiYbO$_2$, the large isotropic thermal parameter of the Li ions suggests this as a possibility. We note here that this distortion is necessarily small and should not significantly affect the $J_1$\---$J_2$ model of the LiYbO$_2$ magnetic lattice. For this reason, analysis of the CEF environment was calculated in the limit assuming only one CEF environment using the $E_i $ = 300 meV data.

\begin{figure}[t]
	\includegraphics[scale=0.275]{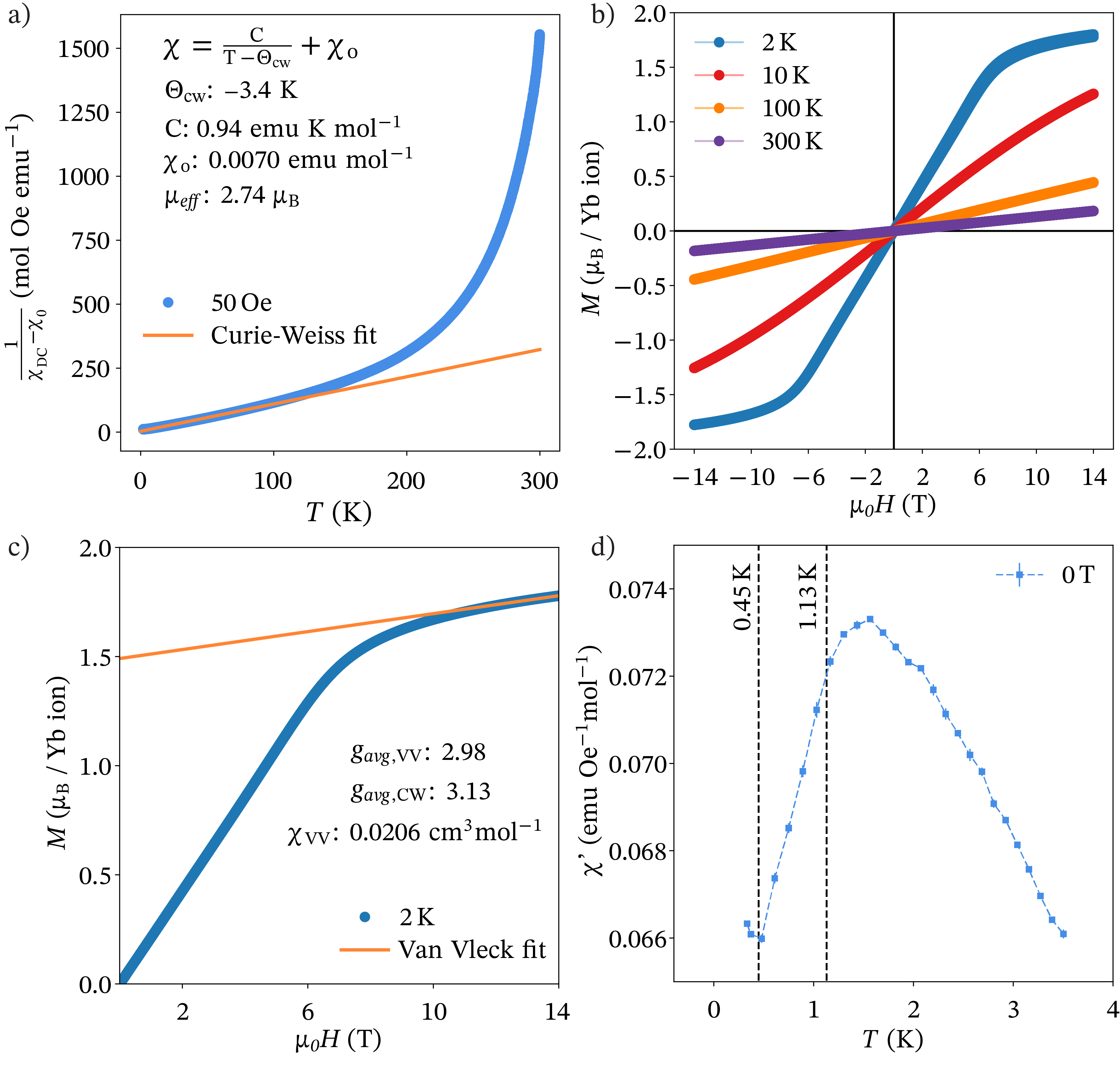}
	\caption{a) Temperature dependence of the inverse magnetic susceptibility of LiYbO$_2$.  Solid line shows the a Curie-Weiss fit to the data between $20<T<100$ K. b) Field dependence of the magnetization collected at a variety of temperatures. c) 2 K isothermal magnetization curve with a linear fit in the saturated state above 10 T. The 0T intercept ($g_{avg}\mu_{B}/2$) provides a powder-averaged $g_{avg,VV}$ and the slope provides $\chi_{VV}$. d) a.c. magnetic susceptibility $\chi `(T)$ data collected for $330$ mK$<T<3.5$ K at zero-field. The two dashes lines at 1.13 K and 0.45 K mark the onset of peaks observed in zero-field heat capacity data.}
	\label{fig:fig3}
\end{figure}

\begin{figure}[]
	\includegraphics[scale=0.45]{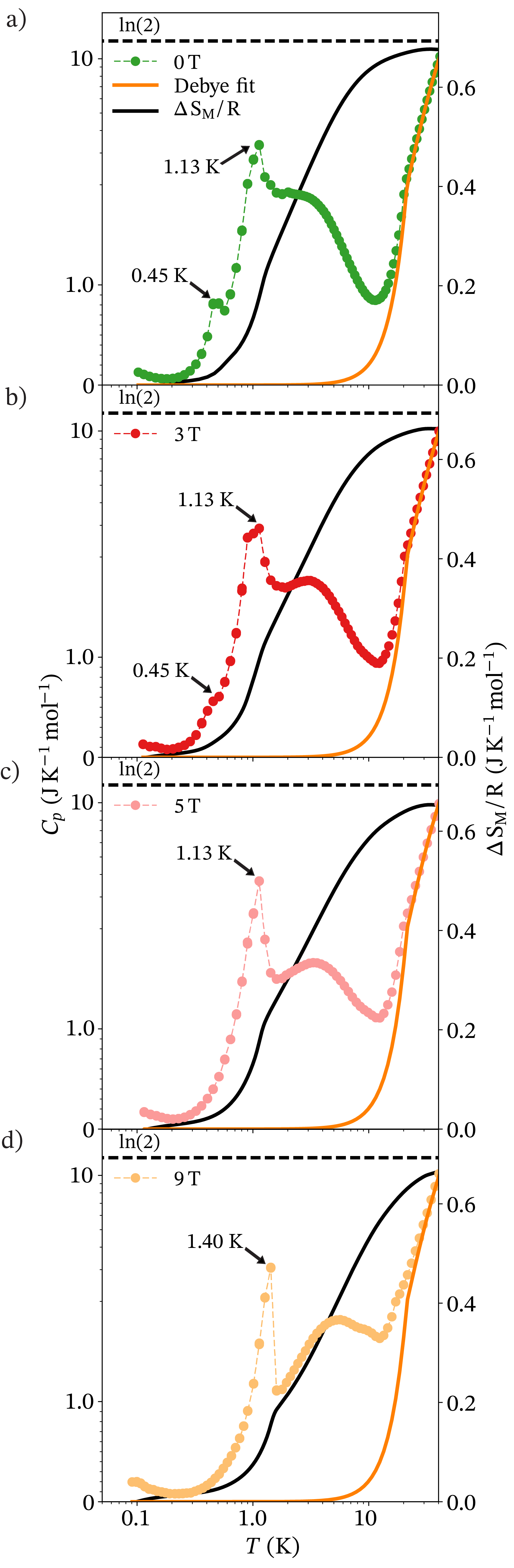}
	\caption{a-d) Specific heat $C(T)$ of LiYbO$_2$ collected as a function of temperature under $\mu_0 H=$ 0, 3, 4, and 9 T. The integrated magnetic entropy $\delta S_M$ is overplotted with the data as a black line. Results from a Debye model of lattice contributions to $C(T)$ are shown as orange lines. The horizontal dashed lines represent $Rln(2)$.}
	\label{fig:fig4}
\end{figure}

\subsection{Magnetization, susceptibility, and heat capacity results}

Figure \ref{fig:fig3} shows the magnetic susceptibility, isothermal magnetization, and a.c. susceptibility measured on powders of LiYbO$_2$. In the low temperature regime where the ground state Kramers doublet is primarily occupied ($T < 100$ K), data were fit to a Curie-Weiss-type behavior with a $\Theta_{CW} = -3.4$ K and an effective moment $\mu_{eff} = 2.74 \ \mu_B$. This implies a powder-averaged $g$-factor $g_{avg,CW} = 3.13$ assuming $J_{eff} = 1/2$ Yb ions. The nonlinearity of the Curie-Weiss fit above 100 K arises due to Van Vleck contributions to the susceptibility that derive from the CEF splitting of the $J = 7/2$ Yb manifold. In order to independently determine $g_{avg}$, the $\chi_{VV}$ contribution to the total susceptibility was fit in the saturated regime ($\mu_0H > 10$ T) of the 2 K isothermal magnetization data shown in Figure \ref{fig:fig3}. In the near-saturated state, the slope of isothermal magnetization yields $\chi_{VV} = 0.0206$ cm$^{3}$ mol$_{Yb}^{-1}$ \cite{li2015rare}, and the intercept of this linear fit with $\mu_0H = 0$ T was utilized to determine the saturated magnetic moment ($g\mu_B/2$) that corresponds to a powder-averaged $g_{avg,VV} = 2.98$. As the Curie-Weiss fit is more susceptible to minor perturbations and background terms, the $g_{avg,VV} = 2.98$ derived from isothermal magnetization data was used for fitting the CEF scheme in Figure \ref{fig:fig5} and Table \ref{tab:tab1}. 

Magnetic susceptibility data in Figure \ref{fig:fig3} explore the low temperature magnetic behavior of LiYbO$_2$.  Two low-temperature ($T < 10$ K) features appear: The first is a broad cusp in susceptibility centered near 1.5 K and is an indication of the likely onset of magnetic correlations. The second feature is a small upturn below 0.45 K. When compared with specific heat measurements in Figure \ref{fig:fig4}, these two features in $\chi^\prime(T)$ coincide with the two sharp anomalies in $C_p(T)$ at $T_{N1}=1.13$ K and $T_{N2}=0.45$ K.  An additional broad peak also appears in $C(T)$ centered near 2 K, likely indicative of the likely onset of short-range correlations. As discussed later in this manuscript, the two lower temperature peaks in $C_p(T)$ mark the staged onset of long-range magnetic order with $T_{N1}$ marking the onset of partial order with disordered relative phases between the $A$ and $B$ Yb-ion sublattices and with $T_{N2}$ marking the onset of complete order between the two sublattices. 

Figure \ref{fig:fig4}a) also displays the total magnetic entropy released upon cooling down to 100 mK. Below 200 mK, a nuclear Schottky feature arises from Yb nuclei as similarly observed in NaYbO$_2$ \cite{bordelon2019field}. Integrating $C_p/T$ between 100 mK and 40 K shows that 98\% of $Rln(2)$ is reached at 0 T, showing that the ordering is complete by 100 mK. Approximately half of $Rln(2)$ is released upon cooling through the broad 2 K peak representing the onset of short range correlations. $C_p(T)$ data were also collected under a series of applied magnetic fields. The onset of $T_{N1}$ stays fixed at 1.13 K from 0 T to 5 T and shifts up to 1.40 K at 9 T. The 0 T heat capacity anomaly at $T_{N2}=0.45$ K begins to broaden at 3 T into a small shoulder of the initial 1.13 K transition and vanishes by 5 T. The broad $C_p(T)$ peak marking the onset of short-range correlations near 2 K shifts to higher temperatures with increasing magnetic field, consistent with a number of other frustrated spin systems \cite{bordelon2019field, li2015rare}. The suppression of the staged $T_{N1}$-$T_{N2}$ ordering under modest magnetic field strengths suggests that zero-field fluctuations/remnant degeneracy likely influence the ordering behavior.

\begin{figure*}[t]
	\includegraphics[width=\textwidth*9/10]{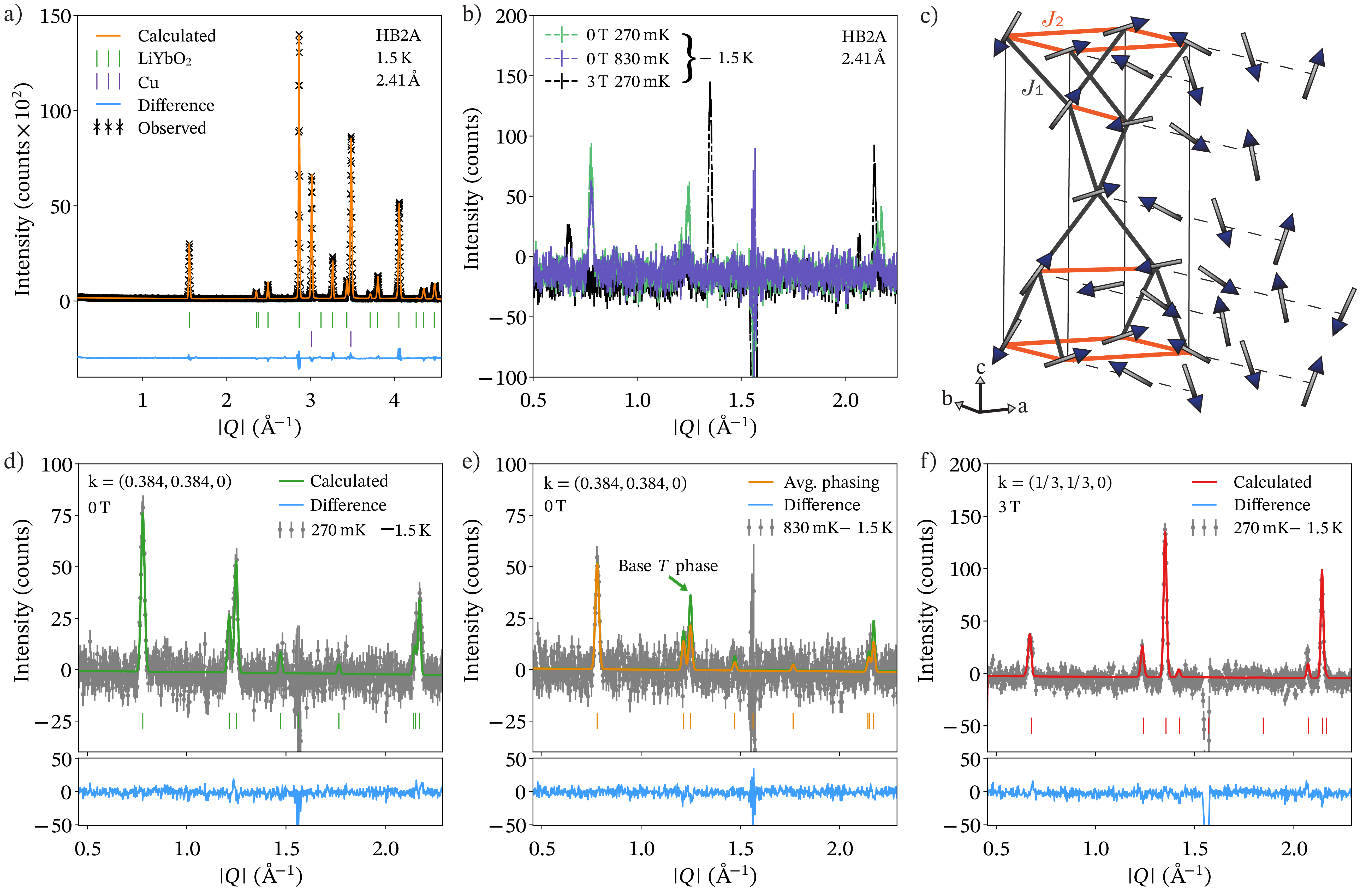}
	\caption{Neutron powder diffraction data collected for LiYbO$_2$ at HB-2A at the High Flux Isotope Reactor. a) Fits to the elastic scattering data at 1.5 K reveal only one structural phase. b) Temperature-subtracted diffraction data ($T-1.5$ K) revealing a series of new magnetic peaks upon cooling. Additionally, at 270 mK and 3 T, another set of magnetic peaks arise. Intensity near 1.5 \AA$^{-1}$ results from slight under/over subtraction of the structural peak at that position in a) and is not a magnetic Bragg reflection. c) Helical magnetic structure fit below the ordering transition $T_{N2}$. d) 270 mK data collected under zero field with the 1.5 K structural data subtracted. Green line shows the resulting fit using the magnetic structure described in the text.  e) 830 mK data collected under zero field with the 1.5 K structural data subtracted.  The orange line shows the partially disordered, intermediate helical state described in the text and the green line shows a fit using the fully ordered helical structure for comparison. f) 270 mK data collected under $\mu H=3$ T with the 1.5 K structural data subtracted.  The red line shows the fit to the commensurate magnetic structure describe in the text.}
	\label{fig:fig6}
\end{figure*}

\begin{table}[]
	\caption{Coefficients of the magnetic basis vectors creating the helical models of the base temperature magnetic structure of LiYbO$_2$ at 0 T and 3 T, where $bv_1 = (100)$, $bv_2 = (010)$, and $bv_3 = (001)$.} 
	\begin{tabular}{c|ccc|ccc}
		\hline
		& \multicolumn{3}{c|}{270 mK, 0 T}           & \multicolumn{3}{c}{270 mK, 3 T}        \\
		& \multicolumn{3}{c|}{${\bf{k}} = (0.384, \pm 0.384, 0)$} & \multicolumn{3}{c}{${\bf{k}} = (1/3, \pm 1/3, 0)$} \\ \hline
		atom ($x$, $y$, $z$)                & $bv_1$       & $bv_2$     & $bv_3$     & $bv_1$       & $bv_2$     & $bv_3$    \\ \hline
		Yb$_1$ (0, 0.75, 0.125) &  0            & -1.26$i$     & 1.26        & 0          & -1.26$i$    & 1.26       \\
		Yb$_2$ (0, 0.25, 0.875) &  0            & -1.26$i$     & 1.26        & 0          & -1.26$i$    & 1.26      
	\end{tabular}
	\label{tab:tab3}
\end{table}

\subsection{Neutron diffraction results}
To further investigate the low-temperature, ordered state, neutron powder diffraction measurements were performed.  Figure \ref{fig:fig6} details the field- and temperature-evolution of magnetic order in LiYbO$_2$ about the $T_{N1}$ and $T_{N2}$ transitions identified in specific heat measurements (Figure \ref{fig:fig4}). Magnetic peaks appear in the powder neutron diffraction data below 1 K, and three regions of ordering were analyzed: (1) In the zero-field low-temperature, fully ordered state ($T<450$ mK); (2) in the zero-field, intermediate ordered state ($450$ mK$<T< 1$ K); and (3) in the field-modified ordered state ($T<450$ mK and $\mu_0 H=3$ T).  Figure \ref{fig:fig6}a) shows the data and structural refinement collected at 1.5 K in the high temperature paramagnetic regime---this is used as nonmagnetic background that is subtracted from the low-temperature data. Figure \ref{fig:fig6}b) shows the subtracted data in each of the above regions overplotted with one another, and each magnetic profile is discussed separately in the following subsections. We note here that in each region, the large difference signal observed slightly above 1.5~\AA$^{-1}$ ~is due to the slight under/over subtraction of a nuclear reflection. 

\subsubsection{Region 1: $\mu_0 H=0$ T , $T<450$ mK}

 At 270 mK, well below $T_{N2}$, a series of peaks appear at incommensurate momentum transfers. These new magnetic reflections are described by a doubly-degenerate ordering wave vector of ${\bf{k}} = (0.384, \pm 0.384, 0)$. The best fit to the data in this regime corresponds to a helical magnetic structure shown in \ref{fig:fig6}c) that is produced from the $\Gamma_1$ irreducible representation (Kovalev scheme) of this space group with the three basis vectors $bv_1 = (1, 0, 0)$, $bv_2 = (0, 1, 0)$, and $bv_3 = (0, 0, 1)$. The helical state is defined by a combination of the ordering wave vector $\bf{k}$ and the helical propagation direction. The latter defines a vector that moments rotate in the plane perpendicular. Best fits for the refinement data were achieved when the helical propagation vector is restricted to the $ab$-plane. However, all helical propagation directions within the $ab$-plane produce equivalent fits to the data.

The fit presented in Figure \ref{fig:fig6}d) corresponds to the instance where helices propagate along the $b$-axis with moments rotating within the $ac$-plane depicted in in Figure \ref{fig:fig6}c). Coefficients of the basis vector representation of this fit are shown in Table \ref{tab:tab3}. Due to the bipartite nature of this lattice, two magnetic Yb$^{3+}$ atoms are defined in the system (denoted as sublattices $A$ and $B$), and in effect, this creates a relative phase difference in the moment rotation between the two sites that is experimentally fit at $0.58\pi$. Additional simulations provided in the Supplemental Material section detail how altering the phasing of the sublattices affects the refinement \cite{supplementalmaterial}. The ordered magnetic moment refined with this fit is $\mu=1.26(10)$ $\mu_B$, comprising 84\% of the expected 1.5 $\mu_B$ moment in a $J_{eff} = 1/2$ system with $g_{avg} = 3$.

\subsubsection{Region 2: $\mu_0 H=0$ T , $450$ mK $<T<1.13$ K}

As the temperature is increased above $T_{N2}$ to 830 mK into the intermediate ordered state, incommensurate magnetic reflections with the same ordering wave vector of ${\bf{k}} = (0.384, \pm 0.384, 0)$ persist (Figure \ref{fig:fig6}e)). Order in this $T_{N1}$ state is seemingly still long-range and the lowest angle reflection can be fit to a Lorentzian peak shape to extract an estimated, minimum correlation length. In both the 270 mK base temperature and 830 mK intermediate temperature regimes, the minimum correlation length corresponds to $\approx$ 364 \AA.  Modeling the pattern of magnetic peaks in this intermediate temperature regime using the same $T_{N2}$ structure as described above however fails to fully capture the data.  As seen in Figure \ref{fig:fig6}e), the $T_{N2}$ (green) structure overestimates reflections near 1.2 \AA$^{-1}$.   

One potential model for the magnetic order in this intermediate temperature regime is to allow the relative phasing of the $A$ and $B$ magnetic sublattices to become disordered upon warming into the $T_{N1}$ state. In other words, helical magnetic order could establish with ${\bf{k}} = (0.384, \pm 0.384, 0)$; however the phasing between Yb-sites would remain disordered prior to selecting a specific phase below $T_{N2}$. This conjecture was modeled by averaging over ten fits using equally-spaced relative phases from zero to $2\pi$ between Yb-sites, and where each fit was calculated using an identical moment size (1.26 $\mu_b$). This averaged phasing model (Figure \ref{fig:fig6}d) orange) captures the relative peak intensities better than the single-phase model used below $T_{N2}$ and is supported by $C(T)$ data showing that additional entropy freezes out below $T_{N2}$. 

\subsubsection{Region 3: $\mu_0 H=3$ T , $T<450$ mK}

Upon applying a magnetic field to the low-temperature ordered state below $T_{N2}$, the magnetic ordering of the system changes. Figure \ref{fig:fig6}f) shows that a $\mu_0 H=3$ T field drives commensurate peaks to appear in place of the incommensurate reflections in the zero-field ordered state. The modified propagation vector corresponds to the doubly-degenerate ${\bf{k}} = (1/3, \pm 1/3, 0)$.  Although the modified ${\bf{k}}$ reflects a locking into a commensurate structure, qualitatively, the details of the ordered state remain similar to the zero-field $T_{N2}$ model. The commensurate 3 T state is still best represented by an $ab$-plane helical magnetic structure with basis vector coefficients displayed in Table \ref{tab:tab3}. The magnetic moment is refined to be $\mu=1.26(9)$ $\mu_B$ and the two Yb-sublattices differ by a relative phase of $0.42\pi$.  

\subsection{Low-energy magnetic fluctuations}

\begin{figure}[t]
	\includegraphics[scale=0.7]{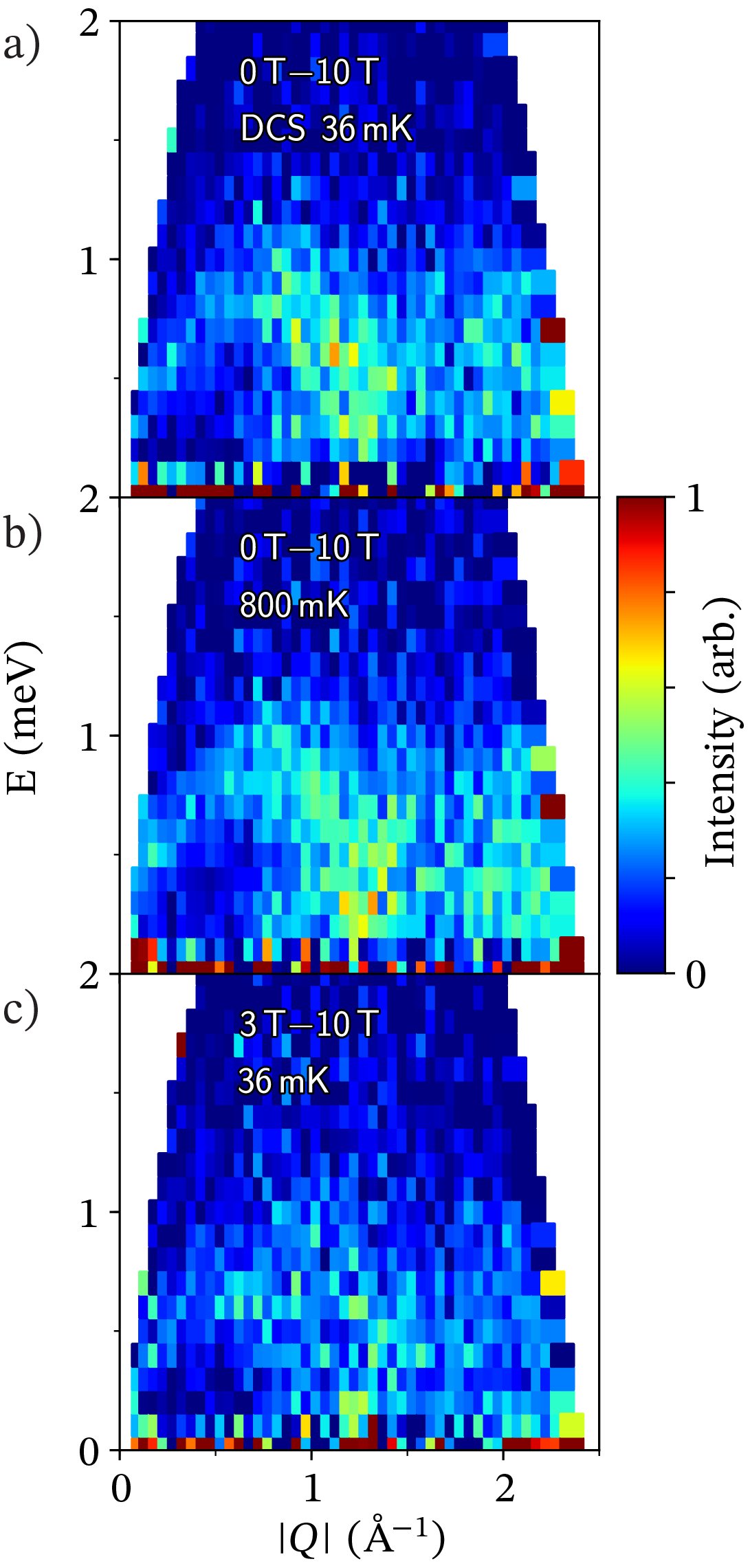}
	\caption{Low-energy inelastic neutron scattering (INS) spectra $S(|Q|,\hbar \omega)$ collected on the DCS spectrometer at a) $\mu_0 H= 0$ T and 36 mK, b)  $\mu_0 H= 0$ T and 800 mK, and c) $\mu_0 H= 3$  T and 36 mK. All data have data collected at 36 mK and 10 T subtracted, where LiYbO$_2$ enters a field-polarized state, indicated by isothermal magnetization data from Figure 4b).}
	\label{fig:fig8}
\end{figure}

The low-energy spin dynamics of Yb moments in LiYbO$_2$ were investigated in all three ordered regimes described in the previous section via inelastic neutron scattering measurements. While the powder-averaged data is difficult to interpret given the complexity of the ordered state, Figure \ref{fig:fig8} plots a series of background-subtracted inelastic spectra that qualitatively illustrate a few key points.  Below $T_{N2}$ and in zero-field, the bandwidth of spin excitations extends to roughly 1 meV. Spectral weight appears to originate from the magnetic zone centers of $k = (q, \pm q, 0)$ (where $q = 0.384$ at 0 T and $q = 1/3$ at 3 T) and the $\Gamma$ point. As the ordered does not change appreciably under moderate fields, the low-energy spectra remain qualitatively similar for both 0 T and 3 T data below $T_{N2}$. Similarly, upon heating from $T_{N2}$ into the $T_{N1}$ state, minimal changes are observed in the inelastic spectra. At 10 T and 36 mK however, LiYbO$_2$ enters a field-polarized state where the low energy spin fluctuations are dramatically suppressed.  The removal of low-energy fluctuations in this high-field data was used to subtract out background contributions in the data shown in Figure \ref{fig:fig8}. There are slight differences in the dynamics of the 0 T and 3 T states in Figure \ref{fig:fig8} that will require future experiments to detail their differences with higher statistics. The raw data for each field and temperature setting are plotted in the Supplemental Material section for reference \cite{supplementalmaterial}.

\section{IV. Theoretical analysis}
In the following subsections, we construct a classical Heisenberg Hamiltonian to describe the interactions of Yb ions in LiYbO$_2$.  We then use this Hamiltonian, extended out to next-nearest neighbors, to model the potential magnetic ground states in LiYbO$_2$ for comparison with experimental data.  Spin excitations are then also modeled in the parameter space predicting magnetic order most closely matching that experimentally observed.  

\subsection{LiYbO$_2$ symmetry analysis}
A minimal Hamiltonian describing the nearest-neighbor (NN) interactions in LiYbO$_2$ ($I4_1/amd$) following symmetry analysis \cite{supplementalmaterial} can be written as
\begin{equation}
\begin{aligned}
	\label{eq:H1}
	H_1=&\sum_{\langle i,j\rangle}J_z S^z_i S^z_j+ J_{xy} (S^x_i S^x_j+ S^y_i S^y_j) \\
	&+ J_\delta (\bm{S}_i\cdot \bm{f}_{ij})(\bm{S}_j\cdot \bm{f}_{ij})
	 	 +J_{cz}(\bm{S}_i\cdot \bm{f}_{ij} S^z_j + S^z_i\cdot \bm{f}_{ij} \hat{\bm{z}}),
	 \end{aligned}
\end{equation}
where $\bm{f}_{ij}$ is the projection of {the bond vector} $\bm{e}_{ij}$ onto the basal plane. The symmetry-allowed next nearest-neighbor (NNN) interactions are written as
\begin{equation}
\begin{aligned}
	\label{eq:H2}
	H_2=&\sum_{\langle\langle i,j\rangle\rangle} J'_z S^z_i S^z_j + J'_{xy}(S^x_iS^x_j+S^y_iS^y_j) \\
	&+ J'_\delta(\bm{S}_i\cdot \bm{e}_{ij})(\bm{S}_j\cdot \bm{e}_{ij}) 
	+ \bm{D}_{ij} \cdot \bm{S}_i\times \bm{S}_j,
\end{aligned}
\end{equation}
where the Dzyaloshinskii-Moriya (DM) vectors for the NNN bonds $\langle ij\rangle$ along $\bm{a}$ and $\bm{b}$ are $\bm{D}_{ij} = (-1)^{\mu(i)} D \bm{a}\times \hat{\bm{z}}$ and $\bm{D}_{ij} =(-1)^{\mu(i)} D\bm{b}\times \hat{\bm{z}}$, respectively. Here $\mu(i)=0,1$ for the sublattice $i=A,B$, respectively, indicating that the sign of the DM vector alternates between layers. 

We hereby restrict our study to the Hamiltonian up to NNN: $H = H_1+H_2$. For $f$-orbital ions such as Yb, the anisotropies $J_\delta$ and $J'_\delta$ are usually negligible, and as a good approximation we take the Heisenberg limit $J_z=J_{xy}=J_1$, and $J'_z=J'_{xy}=J_2$ (see \cite{supplementalmaterial} for a discussion on the effect of $J_z\neq J_{xy}$  and $J'_z\neq J'_{xy}$). This generates as a physical model the $J_1$\---$J_2$ Heisenberg Hamiltonian

\begin{equation}
	\label{eq:Htot}
	H = J_1\sum_{\langle ij\rangle} \bm{S}_i\cdot \bm{S}_j + J_2 \sum_{\langle \langle ij\rangle\rangle} \bm{S}_i \cdot \bm{S}_j + \bm{D}_{ij}\cdot \bm{S}_i\times \bm{S}_j.
\end{equation}

\subsection{The $\bf{J_1}$\---$\bf{J_2}$ model and spiral order}

We first look at the $J_1$\---$J_2$ Heisenberg model on the stretched diamond lattice without the DM term. The classical ground state of this model can be solved exactly. In momentum space, the $J_1$\---$J_2$ Heisenberg model is written as
\begin{equation}\label{shs}
H = \sum_{\bm{q},\mu,\nu} \bm{S}_{\bm{q},\mu} J^{\mu\nu}_{\bm{q}} \bm{S}_{-\bm{q},\nu},
\end{equation}
with
\begin{equation}
\nonumber
\begin{aligned}
J^{11}_{\bm{q}} &= J^{22}_{\bm{q}} = J_2 (\cos \bm{q}\cdot \bm{a} +  \cos \bm{q}\cdot \bm{b}),\\
J^{12}_{\bm{q}} &= J^{21*}_{\bm{q}}=J_1 \left(e^{-i\frac{\bm{q}\cdot \bm{c}}{4}}\cos \frac{\bm{q}\cdot \bm{a}}{2}+e^{i\frac{\bm{q}\cdot \bm{c}}{4}}\cos \frac{\bm{q}\cdot \bm{b}}{2}\right).
\end{aligned}
\end{equation}
Therefore the lower branch of the band is
\begin{equation}
\lambda_{\bm{q}} = J^{11}_{\bm{q}} - | J^{12}_{\bm{q}}|.
\end{equation}
Solving for the minimum of $\lambda_{\bm{q}}$, the classical ground state is an incommensurate spiral, with wave vector 
\begin{equation}
\bm{q} = \frac{2\pi}{a}(q,q,0)\quad \text{or}\quad  \bm{q}= \frac{2\pi}{a}(q,-q,0),
\end{equation}

\begin{figure}[!t]
	\centering
	\includegraphics[width=0.4\textwidth]{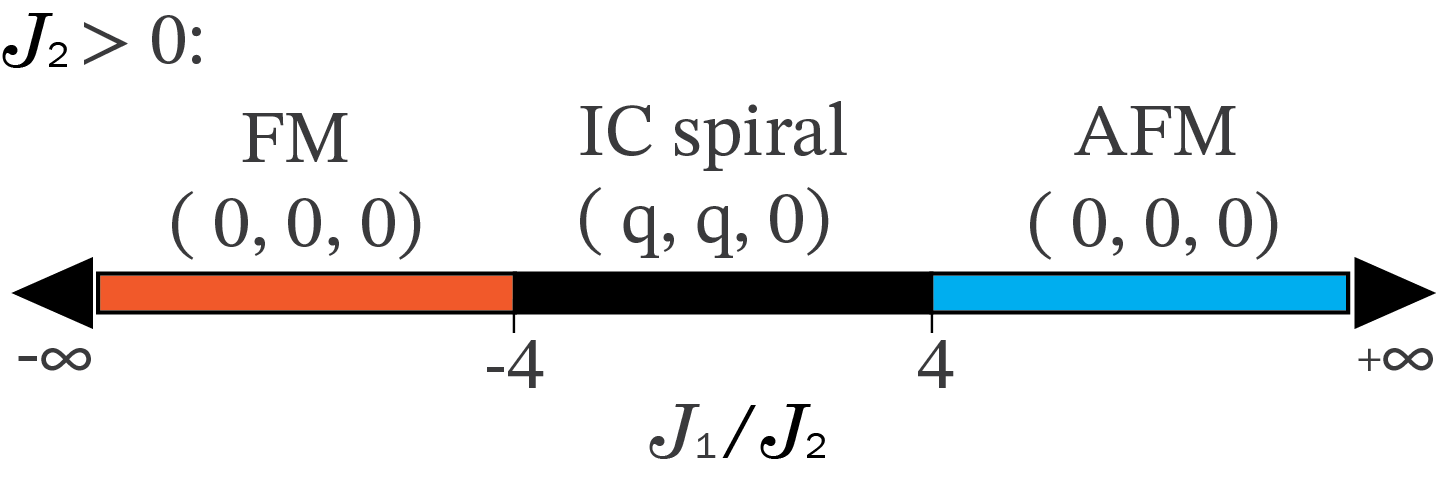}
	\caption{Phase diagram of magnetic order in the $J_1$\---$J_2$ Heisenberg model, assuming $J_2>0$, where ferromagnetic (FM), incommensurate (IC) spiral, and antiferromagnetic (AFM) N\'eel order exist.}\label{j1j2cartoon}
\end{figure}

where
\begin{equation}
\nonumber
q \equiv \left\{\begin{array}{ll} \pm \frac{1}{\pi} \arccos \frac{|J_1|}{4J_2},\\
0,\end{array}\right.
\end{equation}
\begin{equation}
\nonumber
\text{respectively for} \left\{\begin{array}{ll} |J_1|\leq 4J_2,\\
|J_1|>4J_2. \end{array} \right.
\end{equation}
Note that due to the sublattice structure, both the FM and AFM N\'eel orders have $q=0$. From now on we assume $J_2>0$ since spiral order can appear only for a positive $J_2$ (Figure \ref{j1j2cartoon}).
The experimental value for the doubly-degenerate spiral wave vector is $\frac{2\pi}{a}(0.384, \pm 0.384,0)$,
which gives
\begin{equation}
J_1 = \pm 4 \cos (0.384\pi)J_2 = \pm 1.426 J_2.
\end{equation}
The eigenvector corresponding to $\lambda_{\bm{q}}$ is $u_{\bm{q}} = \frac{1}{\sqrt{2}} (e^{i \phi_{\bm{q}}},1)^T$, where the phase $\phi_{\bm{q}} = \pi + \mathrm{Arg} J^{12}_{\bm{q}}$ determines the relative angle or phase between the spins of the two sublattices. The magnetic order then is 
\begin{equation}
\label{mag}
\bm{S}_{\bm{r}_i} = \left(0, \cos \bm{q}\cdot \bm{r}_i, \sin \bm{q}\cdot \bm{r}_i\right) 
\end{equation}
or any coplanar configuration that is related to Eq.~\eqref{mag} by a global SO(3) rotation.

\begin{figure}[!b]
\centering
\includegraphics[width=0.19\textwidth]{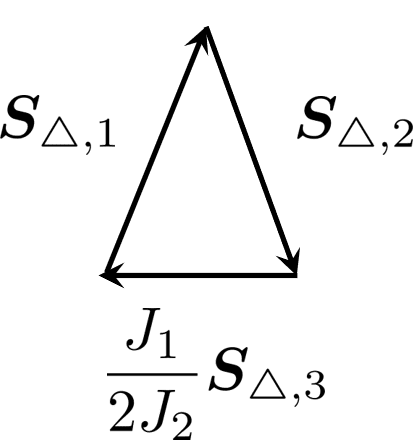}
\caption{The classical ground state condition $\bm{S}_{\triangle,1} + \bm{S}_{\triangle,2} + \frac{J_1}{2J_2}\bm{S}_{\triangle,3}=0$.}\label{triangle}
\end{figure}

A more intuitive, geometrical way to obtain the ground state of the Heisenberg $J_1$\---$J_2$ Hamiltonian is to rewrite it as the sum over all the ``elementary'' triangles $\triangle$ that are enclosed by two NN bonds and one NNN bond, where each NNN bond belongs to only one ``elementary'' triangle while each NN bond is shared by two ``elementary'' triangles. Concretely, for each $\triangle$, label the two spins connected with an NNN bond as $\bm{S}_{\triangle,1}$ and $\bm{S}_{\triangle,2}$, and the third spin as $\bm{S}_{\triangle,3}$, we then have:
\begin{equation}
H = \text{Constant} + \frac{J_2}{2} \sum_{\triangle} \left(\bm{S}_{\triangle,1} + \bm{S}_{\triangle,2} + \frac{J_1}{2J_2}\bm{S}_{\triangle,3}\right)^2.
\end{equation}
Written in this way, the classical ground state is the spin configuration that satisfies $\bm{S}_{\triangle,1} + \bm{S}_{\triangle,2} + \frac{J_1}{2J_2}\bm{S}_{\triangle,3}=0$ for all $\triangle$. Denote the (orientationless) angle between two vectors $\bm{S}_1$ and $\bm{S}_2$ by $\langle \bm{S}_1,\bm{S}_2\rangle$. One easily infers from Figure \ref{triangle} that
\begin{equation}
\begin{aligned}
&	\langle \bm{S}_{\triangle,1},\bm{S}_{\triangle,3}\rangle = \langle \bm{S}_{\triangle,2},\bm{S}_{\triangle,3}\rangle \\
&=
	 \left\{\begin{array}{ll}\pi - \arccos \frac{J_1}{4 J_2}>\frac{\pi}{2},& 4J_2\geq J_1>0 \\ \arccos \frac{|J_1|}{4 J_2}<\frac{\pi}{2},& 4J_2\geq -J_1 >0\end{array}\right.,\\
&\langle \bm{S}_{\triangle,1},\bm{S}_{\triangle,2}\rangle = 2\arccos \frac{|J_1|}{4J_2}.
	 \end{aligned}
 \end{equation}
This result agrees with the exact diagonalization result above. When $J_1 = 1.426 J_2>0$ with a sublattice phasing of $\pi$, the angle between the two spins in a primitive cell is expected to be $\pi - \arccos(1.426/4) = 1.935 \sim 111^\circ$.

\subsection{Effect of other terms; phasing and lattice distortion}

The $J_1$\---$J_2$ model reproduces the spiral phase and the incommensurate wave vector in the ground state of LiYbO$_2$. The angle difference between the nearest spins ($111^\circ$), however, does not agree with the best experimental fitting (staggered in alternating $34^\circ$ and $172^\circ$ angles). One plausible explanation is a small lattice distortion that is outside of resolution of the neutron powder diffraction data.

In this subsection, we study the effect of a lattice distortion on the magnetic order. We assume a simple scenario in which the lattice distortion results in a displacement between two sublattices: suppose the $\mu=1$ sublattice, originally $\delta = \bm{a}/2 + \bm{c}/4$ part from the $\mu=0$ sublattice, is offset by $\bm{\varepsilon}$ from the original position, where $\bm{\varepsilon} = (\epsilon,\epsilon,0)$. In this case the NN vectors from the Yb ion at the origin become $\frac{\bm{a}}{2} +  \frac{\bm{c}}{4} + \bm{\varepsilon}$, $-\frac{\bm{a}}{2} +  \frac{\bm{c}}{4} + \bm{\varepsilon}$, $\frac{\bm{b}}{2} -  \frac{\bm{c}}{4} + \bm{\varepsilon}$, and $-\frac{\bm{b}}{2} -  \frac{\bm{c}}{4} + \bm{\varepsilon}$, which correspond to $J'_1,J''_1,J'_1,J''_1$, respectively. Here we assume antiferromagnetic exchange $J'_1,J''_1>0$ in order to agree with experiment. We can again write down the Hamiltonian in momentum space in the form of Eq.~\eqref{shs}, with modified off-diagonal element
\begin{equation}
\begin{aligned}
J^{12}_{\bm{q}} = J^{21*}_{\bm{q}}&=\frac{J'_1}{2}\left( e^{i \bm{q}\cdot\left(\frac{\bm{a}}{2} +  \frac{\bm{c}}{4} + \bm{\varepsilon}\right)} + e^{i \bm{q}\cdot\left(\frac{\bm{b}}{2} -  \frac{\bm{c}}{4} + \bm{\varepsilon}\right)}\right)\\
&+\frac{J''_1}{2}\left( e^{i \bm{q}\cdot\left(-\frac{\bm{a}}{2} +  \frac{\bm{c}}{4} + \bm{\varepsilon}\right)} + e^{i \bm{q}\cdot\left(-\frac{\bm{b}}{2} -  \frac{\bm{c}}{4} + \bm{\varepsilon}\right)}\right)\\
&=\frac{1}{2} e^{i \bm{q}\cdot \bm{\varepsilon}}\left(J'_1 e^{i \bm{q}\cdot\left(\frac{\bm{a}}{2} +  \frac{\bm{c}}{4}\right)} + J'_1 e^{i \bm{q}\cdot\left(\frac{\bm{b}}{2} -  \frac{\bm{c}}{4}\right)}\right.\\
&\left.+ J''_1 e^{i \bm{q}\cdot\left(-\frac{\bm{a}}{2} +  \frac{\bm{c}}{4}\right)} + J''_1e^{i \bm{q}\cdot\left(-\frac{\bm{b}}{2} -  \frac{\bm{c}}{4}\right)}\right),
\end{aligned}
\end{equation}
where we denote $q_x = \bm{q}\cdot \bm{a}$, $q_y = \bm{q}\cdot \bm{b}$, and $q_z = \bm{q} \cdot \bm{c}$.  It is easy to show that 
\begin{equation}
\begin{aligned}
\lambda_{\bm{q}} \geq J_2 ( \cos q_x + \cos q_y)&-\sqrt{\frac{J'^2_1}{4} + \frac{J''^2_1}{4} +  \frac{1}{2}J'_1 J''_1 \cos q_x}\\
& - \sqrt{\frac{J'^2_1}{4} + \frac{J''^2_1}{4} +  \frac{1}{2}J'_1 J''_1 \cos q_y},
\end{aligned}
\end{equation}
hence the energy minimum is reached at $q_x = q_y \equiv q_0 $ and $q_z = 0$. Here $q_0=0.384\times 2\pi$ is the required experimental value to minimize $f(q) = J_2 \cos q - \sqrt{ \frac{J'^2_1}{2} + \frac{J''^2_1}{2} + J'_1 J''_1 \cos q}$, and we get
$$ \cos q_0 = \frac{J'^2_1 J''^2_1 - 4 J_2^2 (J'^2_1 + J''^2_1)}{8 J_2^2 J'_1J''_1},$$
This equation restricts the value between $J'_1/J_2$ and $J''_1/J_2$. Setting $J'_1 = J''_1=J_1$ recovers the previous undistorted result, $J_1 = 4 \cos\frac{q_0}{2} = 4 \cos \pi q$. The eigenvector corresponding to $\lambda_{\bm{q}}$ is again $u_{\bm{q}} = \frac{1}{\sqrt{2}} (e^{i \phi_{\bm{q}}},1)^T$, where we now have 
\begin{equation}\label{eq:phase}
\begin{aligned}
\phi_{\bm{q}_0} &= \pi +\bm{q}_0\cdot \bm{\varepsilon} + \arctan \left( \tan \left( \frac{\pi}{4} - \beta\right) \tan \frac {q_0}{2}\right)\\
& \approx \pi + \arctan \left( \tan \left( \frac{\pi}{4} - \beta\right) \tan \frac {q_0}{2}\right),
\end{aligned}
\end{equation}
and we define $\tan \beta = J''_1 /J'_1$. The term $\bm{q}\cdot \bm{\varepsilon}$ is small and can be ignored. Eq.~\eqref{eq:phase} suggests that the angle difference between NN spins (which is $\phi_{\bm{q}_0}+ q_0/2$) depends on the spiral wave vector and the ratio of NN bond exchange energies. If we plug in $\phi_{\bm{q}_0} = 360^{\circ}-34^{\circ} = 172^{\circ}$, then we get $\tan \beta \approx 6$. This means that in our simple lattice distortion scenario, a large exchange ratio is needed in order to reproduce the experimentally observed order. 

We note that the DM contribution vanishes if different layers are assumed to have the same order: assume $D\ll J_1,J_2$; suppose the coplanar order is normal to $\bm{n}$, then the DM interaction in layer $l$ is proportional to $(-1)^{\mu(l)}D (\bm{a}-\bm{b}) \cdot \bm{n} \sin qa$. The sign $(-1)^{\mu(l)}$ indicates that neighboring layers (belonging to different sublattices $A$ and $B$) have opposite contributions, leading to a vanshing DM energy.

\subsection{Linear spin wave theory}

In this subsection, we present simulations of the dynamical structure factor using linear spin wave theory. An undistorted lattice is assumed. Introducing Holstein-Primakoff (HP) bosons
\begin{equation}
\bm{S}_i \cdot \bm{a}_i = \sqrt{s}\frac{a_i+a^\dag_i}{\sqrt{2}},\,\,
\bm{S}_i \cdot \bm{b}_i = \sqrt{s} \frac{a_i - a^\dag_i}{\sqrt{2}i},\,\,
\bm{S}_i \cdot \bm{c}_i = s - n_i
\end{equation}
where $\bm{c}_i = \bm{u}\cos \widetilde{\bm{q}}\cdot \bm{r}_i + \bm{v} \sin \widetilde{\bm{q}}\cdot \bm{r}_i$ is the spin order $(\bm{u}$ and $\bm{v}$ are orthogonal unit vectors spanning the order plane), $\bm{b}_i = \bm{u}\times \bm{v}$, and $\bm{a}_i = \bm{b}_i\times \bm{c}_i$. We define $\widetilde{\bm{q}} = \frac{2\pi}{a}(1-q,1-q,0)$ to remind that the angle between NN spins is obtuse in the $J_1$\---$J_2$ model. The spin wave Hamiltonian is then
\begin{equation}
H = \sum_{\bm{k}\in \text{BZ}^+}\Phi^\dag_{\bm{k}}\mathcal{H}(\bm{k}) \Phi_{\bm{k}},
\end{equation}
where $\Phi_{\bm{k}} = \left(a_{\bm{k},0}, a_{\bm{k},1},a^\dag_{-\bm{k},0}, a^\dag_{-\bm{k},1}\right)^T$ are the HP bosons in momentum space, and
\begin{equation}\label{hs}
\mathcal{H}(\bm{k})
=
2\left(\begin{array}{cccc}
h_{11} &h_{12}&p_{11}& p_{12}\\
h_{12}^*& h_{11} &p_{12}^*&p_{11}\\
p_{11}&p_{12}& h_{11} &h_{12}\\
p_{12}^*& p_{11}& h_{12}^*& h_{11}
\end{array}\right),
\end{equation}
with
\begin{subequations}\label{entries}
\begin{eqnarray}
h_{11} &=& J_2\sum_{\bm{\delta} = \bm{a},\bm{b}} \left(2s\cos \bm{k}\cdot \bm{\delta} \left[\frac{1}{4}(c_{\bm{\delta}}+1)\right]-s c_{\bm{\delta}}\right)\notag\\
&& - J_1\sum_{\bm{\delta} = \pm\frac{\bm{a}}{2}-\frac{\bm{c}}{4},\pm\frac{\bm{b}}{2}+\frac{\bm{c}}{4}} \frac{s}{2} c_{\bm{\delta}},\\
h_{12} &=& J_1\sum_{\bm{\delta} = \pm\frac{\bm{a}}{2}-\frac{\bm{c}}{4},\pm\frac{\bm{b}}{2}+\frac{\bm{c}}{4}} s e^{i\bm{k}\cdot \bm{\delta}} \left[\frac{1}{4}(c_{\bm{\delta}}+1)\right],\\
p_{11} &=& J_2\sum_{\bm{\delta} = \bm{a},\bm{b}} 2s \cos \bm{k}\cdot \bm{\delta} \left[\frac{1}{4}(c_{\bm{\delta}}-1)\right],\\
p_{12} &=&  J_1\sum_{\bm{\delta} = \pm\frac{\bm{a}}{2}-\frac{\bm{c}}{4},\pm\frac{\bm{b}}{2}+\frac{\bm{c}}{4}} s e^{i \bm{k}\cdot \bm{\delta}} \left[\frac{1}{4}(c_{\bm{\delta}}-1)\right],
\end{eqnarray}
\end{subequations}
where we defined
$$c_{\bm{\delta}} \equiv \cos \widetilde{\bm{q}}\cdot \bm{\delta} = \left\{\begin{array}{ll}- J_1/4J_2,& \bm{\delta} \in \text{NN},\\
2 \left(\frac{J_1}{4 J_2}\right)^2 - 1,& \bm{\delta} \in \text{NNN}.\end{array}\right.$$

The boson canonical commutation relation is preserved by the diagonalization $V^\dag_{\bm{k}} \mathcal{H}(\bm{k}) V_{\bm{k}} = \Lambda_{\bm{k}}$, $\Phi_{\bm{k}} = V_{\bm{k}} \Psi_{\bm{k}}$, where $V^\dag_{\bm{k}}J V_{\bm{k}} = J\equiv\mathrm{Diag}(1,1,-1,-1)$. Diagonalizing $J\mathcal{H}(\bm{k})$ then gives the spin wave spectrum $\Lambda=(\lambda_1,\lambda_2,-\lambda_1,-\lambda_2)$, with
\begin{equation}\label{sp}
\lambda_{1,2} = \sqrt{(h_{11} \pm |h_{12}|)^2 - (p_{11} \mp |q_{12}|)^2}.
\end{equation}

\begin{figure}
	\centering
	\includegraphics[width=0.4\textwidth]{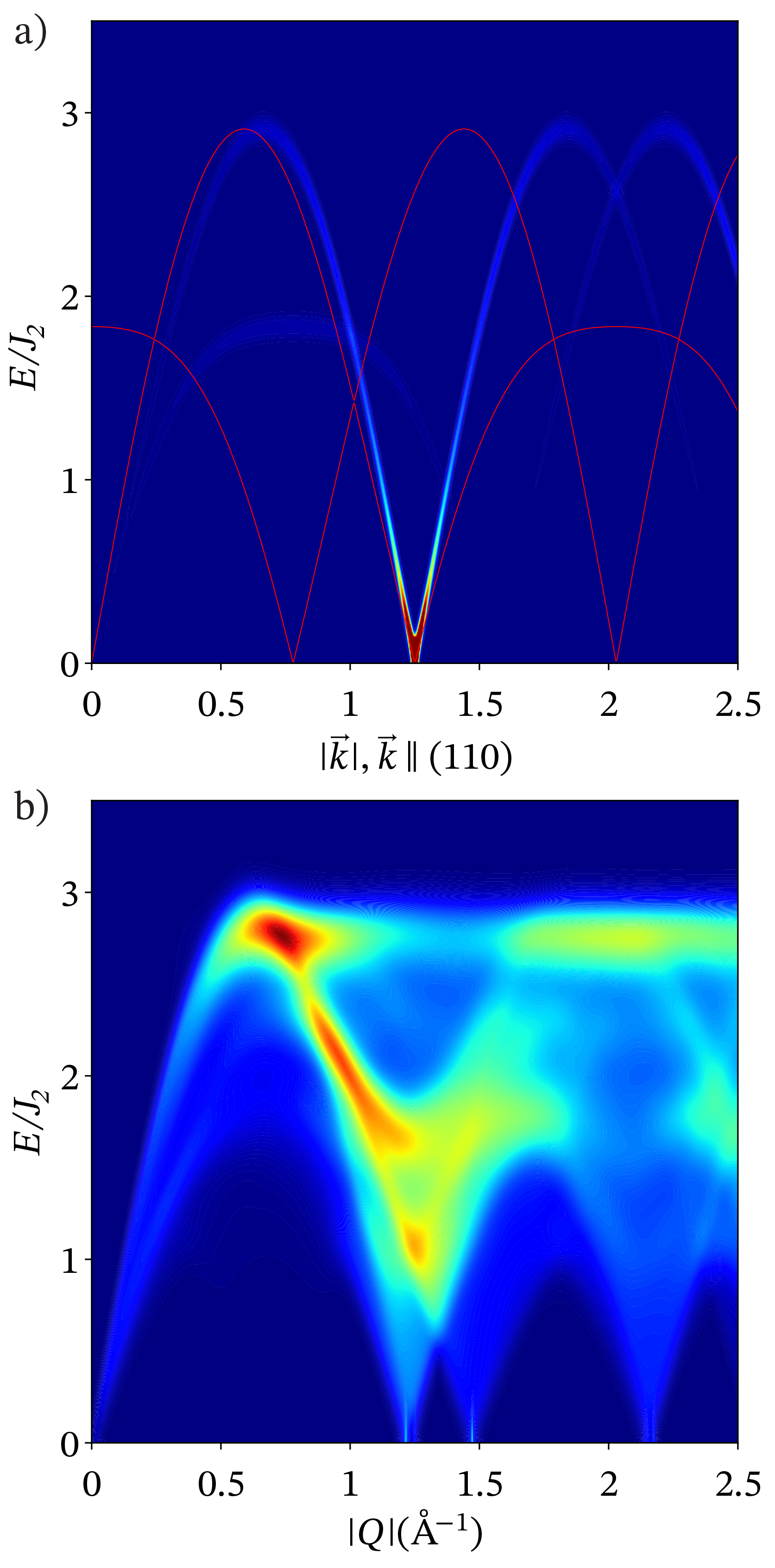}
	\caption{a) Spin wave spectrum (red lines) and the structure factor simulation for $J_1=1.42565 J_2>0$. Both along the (110) direction. b) Angular averaged structure factor for $J_1=1.42565 J_2>0$.}
	\label{spectrum_110}
\end{figure}

The spin wave spectrum \eqref{sp} along the (110) direction is shown in Figure \ref{spectrum_110}a. One observes that the spectrum is gapless at
\begin{equation}
\bm{q} = (0,0,0),\quad \pm\frac{2\pi}{a}(q,q,0),\quad \text{and}\quad \pm \frac{2\pi}{a}(1-q,1-q,0),
\end{equation}

and the momenta that are related to $\bm{q}$ by a $C_4$ rotation along $(001)$ or translation by reciprocal lattice vectors.

We then derive an expression for the dynamical structure factor, which is the Fourier transform of spin-spin correlation function. One obtains
\begin{equation}\label{skw}
\begin{aligned}
&\mathcal{S}(\bm{k},\omega)\\
&=\sum_{i,j=1}^3(\delta_{ij} -(\hat{\bm{k}})_i(\hat{\bm{k}})_j)
\sum_{\mu,\nu=0}^1\langle m^i_\mu(-\bm{k},-\omega)m^j_\nu(\bm{k},\omega)\rangle\\
&= 2s \mu^2_B \sum_{e=1}^4
\delta(\omega - J \lambda_{\bm{k}-\widetilde{\bm{q}},e})
\left[V_{\bm{k}-\widetilde{\bm{q}}}^\dag {K}_1^\dag g^\dag P_{\bm{k}} g {K}_1 V_{\bm{k}-\widetilde{\bm{q}}}\right]_{e,e} \\
&\qquad\qquad + \delta(\omega - J \lambda_{\bm{k}+\widetilde{\bm{q}},e})
\left[V_{\bm{k}+\widetilde{\bm{q}}}^\dag {K}^T_1 g^\dag P_{\bm{k}} g {K}^*_1 V_{\bm{k}+\widetilde{\bm{q}}}\right]_{e,e}\\
& \qquad\qquad + \delta(\omega - J \lambda_{\bm{k},e})  \left[V^\dag_{\bm{k}} {K}_2^\dag g^\dag P_{\bm{k}}g {K}_2 V_{\bm{k}}\right]_{e,e},
\end{aligned}
\end{equation}
where we defined projector $P_{\bm{k}} = 1_{3\times 3}- \hat{\bm{k}}\hat{\bm{k}}^T$.  The derivation and the notation for ${K}_{1,2}$ and $g$ can be found in the Supplementary Material \cite{supplementalmaterial}. From Eq.~\eqref{skw}, it is clear that the structure factor intensity at one $\bm{k}$ receives contributions from three momenta: $\bm{k}\pm \widetilde{\bm{q}}$ and $\bm{k}$. The simulated structure factor according to Eq.~\eqref{skw} is shown in Figure \ref{spectrum_110}a) for a specific $(1,1,0)$ direction, and in Figure \ref{spectrum_110}b) for the angular averaged result.  One of the main features at low-energy is the vanishing intensity at $\Gamma$ and $|\bm{q}| = \frac{2\sqrt{2}\pi}{a}0.384$, where the spin wave spectrum is gapless, and one would naively expect a strong intensity peak at zero energy due to singular BdG Hamiltonian at these momenta. Physically the ``missing'' intensity is a consequence of the destructive interference of the two sublattices at $\Gamma$ and $\bm{q}$ that leads to vanishing contribution to the structure factor. The same interference pattern is also true for the static structure factor. The perfect cancellation is really a consequence of the (undistorted) $J_1$\---$J_2$ Heisenberg model. On the other hand, the persistence of high intensities at $\Gamma$ and $\bm{q}$ from the neutron experiment suggests this cancellation is partially lifted in the real material due to other effects not captured by the $J_1$\---$J_2$ Heisenberg model.

\subsection{Free energy analysis}

The classical ground state of the $J_1$\---$J_2$ Heisenberg model has a global SO(3) symmetry due to the freedom in choosing the spiral plane. Since the lattice only has discrete symmetries, it is likely that this continuous symmetry is lifted due to other effects, such as spin-orbit coupling and fluctuations, and it is the goal of this section to address this issue energetically from a symmetry point of view. Specifically, we will examine the symmetry constraints on the free energy. We first write down the spiral order parameter. Assuming the spiral plane is spanned by two orthogonal vectors $\bm{u}$ and $\bm{v}$, the order parameter can be chosen as the Fourier transform of the magnetic order, which can be written as

\begin{equation}
\bm{d} = e^{i \theta(\bm{r})} (l\bm{u} + im \bm{v}),
\end{equation}
where $\theta(\bm{r})$ determines the direction of the spins in the spiral plane. While it is a constant in the spiral phase, spatial fluctuation of $\theta$ must be considered near the incommensurate-to commensurate (IC-C) transition. Note we have introduced $l$ and $m$ to account for either perfect circular ($l=m$, no net magnetization), elliptical ($m\neq l >0$) or linear ($m=0$) polarization, which correspond to zero, low and high magnetic fields, respectively.

We first look at the zero-field case, $l=m$. Following Lee and Balents \cite{lee-balents}, we seek to write down the free energy for the order parameter to quadratic order using symmetry considerations. Out of the symmetry generators $T_{1,2,3}$, $S_{4z}$, $C_{2y}$ and $P$, the little group of the wave vector $\widetilde{\bm{q}}$ contains $P$, $T_{1,2,3}$, $S_{4z}^2$, and $S^3_{4z}C_{2y}\colon (x,y,z)\rightarrow (y-1/2,x-1/2,3/2-z)$. Under these symmetries, the order parameter transforms as

\begin{subequations}\label{pstts}
\begin{eqnarray}
P\colon&& \bm{d}\rightarrow e^{i\pi \widetilde{q}} \bm{d}^*,\\
S_{4z}^2\colon&& \bm{d}\rightarrow \mathrm{Diag}(-1,-1,1)\bm{d}^*,\\
T_{1,2}\colon&& \bm{d}\rightarrow \bm{d},\\
T_3\colon&& \bm{d}\rightarrow e^{- 2i \pi \widetilde{q}} \bm{d},\\
S^3_{4z}C_{2y}\colon&& \bm{d}\rightarrow \left(\begin{array}{ccc} &1 &\\1 &&\\ && 1\end{array}\right)e^{i 2\pi\widetilde{q}} \bm{d},
\end{eqnarray}
\end{subequations}
where the last symmetry operation can be composed with $T_3$ to get $T_3S^3_{4z}C_{2y}\colon \bm{d}\rightarrow (d_y,d_x,d_z)$. From this, one can write down a free energy density that is quadratic in $\bm{d}$:
\begin{equation}\label{f110}
f(\bm{d}) = c_0 |\bm{d}|^2 + c_1 (d^*_1d_2+c.c.)+c_2 d^*_3 d_3.
\end{equation}
By minimizing this free energy one finds there are three choices for the spiral plane depending on the value of $c_1$ and $c_2$ \cite{lee-balents}: the normal of the order plane can be along either $(001)$, $(1\bar{1}0)$, or $(110)$. 

The result above applies to a generally incommensurate wave vector $\widetilde{q}$ at zero magnetic field. As the field is switched on, the spiral order ceases to be circularly polarized, and the unequal components $l\neq m$ allow for nonzero net magnetization. As a consequence, some of the symmetry transformations in \eqref{pstts} are no longer valid and need to be modified. Nevertheless, we assume that all the symmetry transformations in \eqref{pstts} remain approximately valid at small field. Under these assumptions, we proceed to an explanation of the IC-C transition at 3\,T. The commensurate phase has a three-unit cell order with corresponding wave vector $\bm{q} = 2\pi(\frac{1}{3},\frac{1}{3},0)$. In this phase, another term can be added to the free energy density:
\begin{equation}\label{fC}
f_{\text{C}} = f(\bm{d}) - \widetilde{c}_6 \left((\bm{d}\cdot \bm{d})^3 + c.c.\right).
\end{equation}
The development of unequal $l$ and $m$ can be further modeled phenomenologically by fourth-order terms in the free energy such as $\beta_2|\bm{d}\cdot \bm{d}|^2 + \chi_1 H^2(\bm{d}^*\cdot \bm{d}) + \chi_2 | \bm{H}\cdot \bm{d}|^2$, which we do not discuss here but instead refer to Ref. \cite{zhitom}.

In the following, we show that the IC-C transition can be described phenomenologically by a sine-Gordon model. For given $J_1$ and $J_2$, assume $\bm{q}$ is the (generally incommensurate) ground state spiral wave vector, while $\bm{k}$ is a nearby commensurate wave vector. Assume $\bm{q} = \bm{k}+\bm{\delta k}+\bm{\nabla} \theta$, where $\bm{\nabla} \theta$ denotes the spatial fluctuation of the order parameter. The classical energy can be expanded around $\bm{k}$:
\begin{equation}
\lambda = \lambda_0 + 2 \bm{\delta}\cdot \bm{\nabla}\theta + \frac{\kappa_{xy}}{2}((\partial_x \theta)^2+(\partial_y \theta)^2) + \frac{\kappa_z}{2} (\partial_z \theta)^2,
\end{equation}
where $\lambda_0 = - \frac{J_1^2}{4J_2}-2J_2$, and the rigidity for $\theta$ is
\begin{equation}
\kappa_{xy} = -\frac{a^2}{16 J_2}(J_1^2-16 J_2^2),\qquad \kappa_z = \frac{c^2J_1^2}{32 J_2}.
\end{equation}
importantly, a term linear in the gradient of $\theta$ exists, with coefficient $\bm{\delta} = \kappa_{xy}\bm{\delta k}$. A full theory for $\theta$ then appears as
\begin{equation}\label{ftf}
F[\theta] = A \int d^3x \left( \frac{\kappa}{2}(\bm{\nabla}\theta)^2 + 2 \bm{\delta}\cdot \bm{\nabla} \theta
- c_6 \cos 6 \theta\right),
\end{equation}
where the last term comes from Eq.~\eqref{fC} with $c_6 \sim (l^2-m^2)^3 c_6$. This is the sine-Gordon model that has been analyzed in numerous works; see e.g. Ref. \cite{zhitom}. The basic physics is that the soliton number $N$ of the lowest energy solution to the free energy functional \eqref{ftf} distinguishes commensurate phase ($N=0$) and incommensurate phase ($N = \pm 1$); the C-IC transition then is determined by the energetics of $N=0$ and $N\neq 0$ configurations, with critical relation $\kappa^2 c_6/4\kappa \delta k = \pi^2/32$ ($\kappa^2 c_6/4\kappa \delta k < \pi^2/32$ gives the incommensurate phase). Since the elliptic polarization is induced by magnetic field, following Ref.~\cite{zhitom} we conclude that the coefficient $c_6\propto (l^2-m^2)^3  \propto H^6$, and that increasing the magnetic field will inevitably induce an IC-C transition.

\section{V. Discussion}

\begin{figure}[t]
	\includegraphics[width=0.45\textwidth]{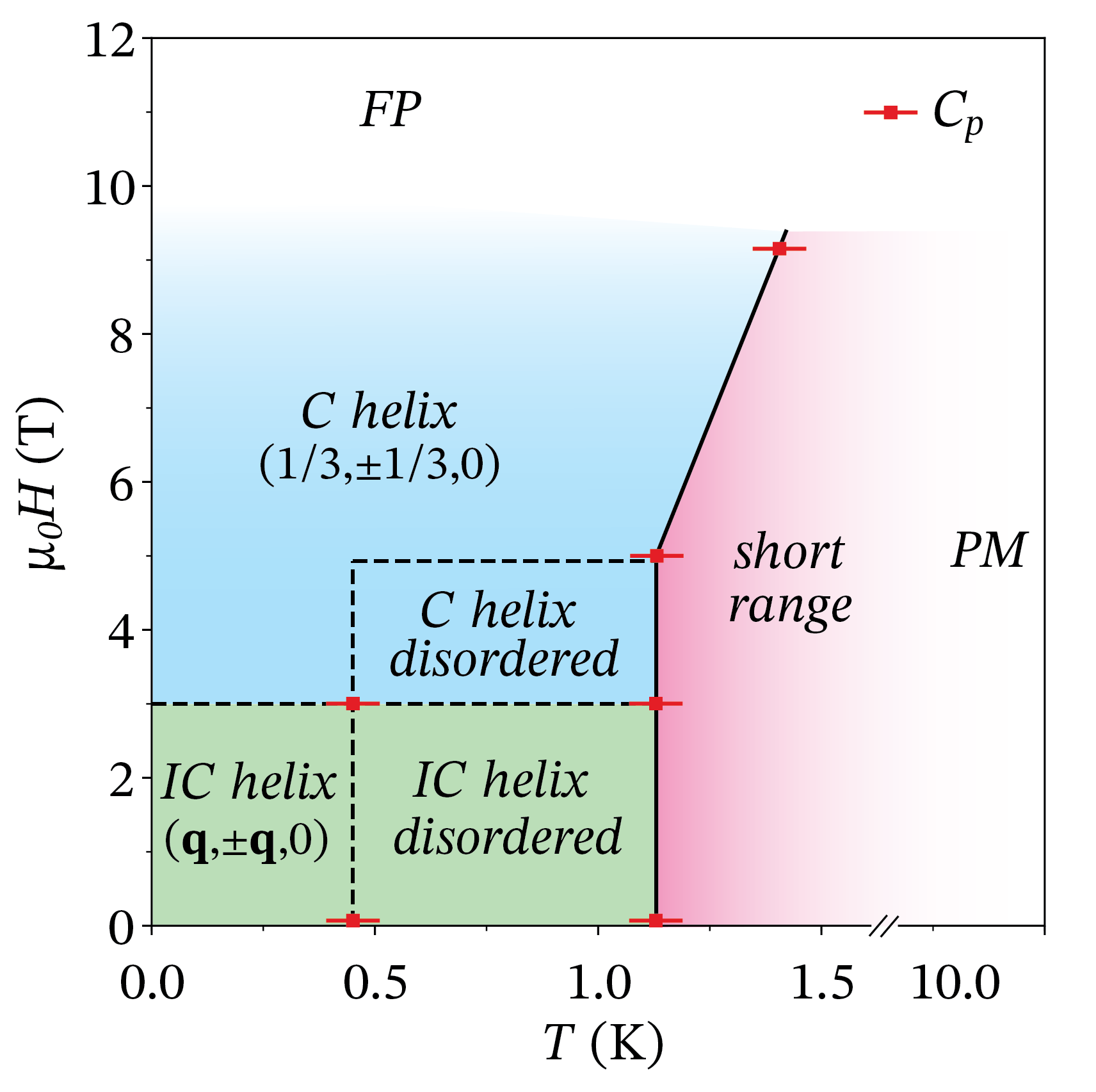}
	\caption{Proposed powder-averaged, low-temperature ($H$, $T$) diagram of LiYbO$_2$ extracted from a combination of specific heat ($C_p$) measurements and elastic neutron powder diffraction data. At high temperature, LiYbO$_2$ is in the paramagnetic ($PM$) phase. Below approximately 10 K, specific heat shows a broad feature where roughly half of the magnetic entropy of $Rln(2)$ is released and signifies the onset of short-range magnetic correlations. A sharp anomaly at 1.13 K at 0, 3, and 5 T and 1.40 K at 9 T in specific heat measurements shows where long-range magnetic order sets in. Combining specific heat data with neutron powder diffraction data suggests that the temperature regime between 0.45 K and 1.13 K consists of a helical magnetic structure with disordered phasing between the two interpenetrating Yb sublattices. The system undergoes a lock-in phase transition from an incommensurate helical structure at zero field to a commensurate structure at 3 T.}
	\label{fig:fig10}
\end{figure}

LiYbO$_2$ shows a rich magnetic phase diagram (see Figure \ref{fig:fig10}) with inherent similarities to the $A$-site transition metal spinels and the $J_1$\---$J_2$ diamond lattice model, indicating that the underlying physics of both systems arises from the same bipartite frustration. The $J_1$\---$J_2$ model on the ideal diamond lattice with $J_2/|J_1| > 1/8$, produces frustrated spiral order with wave vectors directed along the high-symmetry directions of the lattice (e.g. $(q, q, q)$, $(q, q, 0)$, $(0, 0, q)$) and simliar spiral order also appears in tetragonaly elongated diamond lattice of LiYbO$_2$ near $|J_1|\leq 4J_2$. Spiral wave vectors in the distorted case are however limited to $(q, \pm q, 0)$, and tetragonal distortion lifts the degeneracy of the spiral spin liquid surface predicted for the perfect diamond lattice \cite{bergman-balents, lee-balents,buessen-trebst}. 

Curiously, in zero-field, the long-range helical ground state forms through two successive magnetic transitions upon cooling. An intermediate state formed upon cooling below $T_{N1}$ is best fit by modeling a spiral state on each Yb-site but with disordered relative phasing between the two spirals.  This apparent frustration in the relative phase between magnetic sublattices and the formation of a partially ordered state is also likely reflected in the departure of the relative phasing between Yb-ions within the fully ordered state (below $T_{N2}$) from the predictions of the Heisenberg $J_1$\---$J_2$ model.  Specifically, the model predicts that moments rotate along all $A$-to-$B$ sublattice bonds equivalently (i.e. the angle difference between every NN spin is $111^\circ$), while the experimental data suggests that moments rotate in a staggered fashion, where the first $A$-to-$B$ sublattice bond is $34^\circ$ and the second is $172^\circ$. This generates a magnetic structure in which pairs of spins between the $A$ and $B$ sublattices are nearly aligned antiparallel. 

While CEF data suggest the presence of two Yb environments in the lattice, this is not readily apparent in the average structural data, suggesting that the distortion responsible for this is reasonably subtle.  Given the large distortion required for the model to produce the experimentally observed phasing between Yb-moments, 
the possible origin for the phase difference instead lies in the presence of anisotropic exchange interactions in LiYbO$_2$. We note however that, assuming spiral order with a single wave vector $q$, including Ising type of anisotropy at NN and NNN level does not help in explaining the disagreement between theory and experiment (further details in Supplementary Materials \cite{supplementalmaterial}). Resolving the possibility of other anisotropic terms in the Hamiltonian as well as the precise nature of the anomalous state between 0.45 K $< T <$ 1.13 K will require future single crystal studies.


The incommensurate helical structure in LiYbO$_2$ evolves into a commensurate helical structure when $\mu_0 H=3$ T is applied. A similar type of ``lock-in'' incommensurate-to-commensurate (IC-C) phase transition occurs in the $A$-site spinels, originating from magnetic anisotropy on top of the $J_1$\---$J_2$ model \cite{lee-balents}. Anisotropy accounts for the change from an incommensurate $(q, \pm q, 0)$ helical phase to a commensurate one in MnSc$_2$S$_4$ \cite{lee-balents, gao-ruegg, iqbal-reuther} and CoCr$_2$O$_4$ \cite{chang2009crossover, lawes2006dielectric, chen2013coexistence} with decreasing temperature. In LiYbO$_2$ however, the field-driven ``lock-in'' phase transition is captured within the sine-Gordon model in Eq.~\eqref{fC} without the need to perturb the Heisenberg $J_1$\---$J_2$ model.

In fact, a considerable amount of the zero-field magnetic behavior of LiYbO$_2$ is captured at the ideal Heisenberg $J_1$\---$J_2$ limit. The doubly-degenerate ordering wave vector $(q, \pm q, 0)$ predicted by the model is reproduced in the fits to elastic neutron diffraction data, and the theory predicts that the spiral structure's ordering plane should be along $(0,0,1)$, $(1,1,0)$, or $(1,\bar{1},0)$. Experimental fits in Figure \ref{fig:fig6} and Table \ref{tab:tab3} rule out the $(0,0,1)$ ordering plane and the remaining planes of $(a,b,0)$ can not be distinguished with the present powder data. Future single crystal neutron experiments could reveal if the ordering plane aligns with the energy minimization in the $(1,1,0)$ or $(1,\bar{1},0)$ planes.

Additionally, the extracted value of $|J_1|/J_2$ = 1.426 from the $J_1$\---$J_2$ model makes intuitive sense within the chemical lattice. It is unsurprising that the two magnetic interactions would be comparable in strength due to their relative superexchange pathways. In comparison, materials such as KRuO$_4$ \cite{marjerrison2016structure} and KOsO$_4$ \cite{song2014unquenched, injac2019structural} share the same $I4_1/amd$ magnetic sublattice comprised of Ru and Os ions, but break the oxygen-based superexchange connection along $J_2$. In these systems, magnetic order resides in the $J_2 = 0$ limit of the Heisenberg $J_1$\---$J_2$ model, where moments order within a Ne\'el antiferromagnetic state and an unfrustrated $J_1$ \cite{marjerrison2016structure, song2014unquenched, injac2019structural}.

Calculations of low-energy spin excitations with the parameters obtained from the $J_1$\---$J_2$ model largely reproduce the low-energy INS spectrum in Figures \ref{fig:fig8} and \ref{spectrum_110} with $J_2 \approx 1/3$ meV and $J_1 \approx 0.475$ meV. One difference appears in the spectral weight at the $\Gamma$ and $|\bm{q}| = \frac{2\sqrt{2}\pi}{a}\times0.384$ positions, where a cancellation of the simulated structure factor intensity occurs due to destructive interference of the two sublattices at these momenta. 
This cancellation does not occur in the experimental data due to the difference in phasing between Yb-moments relative to the predictions of the $J_1$\---$J_2$ model .

Despite this minor deviation, rooted in the relative phasing between the Yb-sublattices, our work establishes that LiYbO$_2$ contains a tetragonally-elongated diamond lattice largely captured by the Heisenberg $J_1$\---$J_2$ model. To the best of our knowledge, reports of diamond lattices decorated with trivalent lanthanide ions are rare, and, based upon our results, we expect that an ideal diamond lattice decorated with Yb$^{3+}$ moments may reside close to the ideal Heisenberg limit. Such an ideal cubic $Ln$-ion diamond lattice would be a promising platform for manifesting (quantum) spiral spin liquid states, similar to transition metal spinels, while potentially avoiding the complications of extended exchange interactions born from $d$-electron systems. 

\section{VI. Conclusions}

LiYbO$_2$ provides an interesting material manifestation of localized
$f$-electron moments decorating a frustrated diamond-like lattice.
Long-range incommensurate spiral magnetic order of
${\bf{k}} = (0.384, \pm 0.384, 0)$ forms in the ground state, which
seemingly manifests through a two-step ordering process via a
partially ordered intermediate state. Upon applying an external
magnetic field, magnetic order becomes commensurate with the lattice
with ${\bf{k}} = (1/3, \pm 1/3, 0)$ through a ``lock-in'' phase
transition. Remarkably, the majority of this behavior in LiYbO$_2$ can
be captured in the Heisenberg $J_1$\---$J_2$ limit where the magnetic
Yb$^{3+}$ ions are split into two interpenetrating $A$-$B$
sublattices. This model was explicitly re-derived and tuned for
LiYbO$_2$, and it is directly related to a physical elongation of the
diamond lattice Heisenberg $J_1$\---$J_2$ model.  Differences in the
relative phasing of $A$-$B$ sublattices between the Heisenberg model
and the observed magnetic structure suggest additional interactions
and quantum effects may be present in LiYbO$_2$.  This is possibly related to
the observation of crystal field splittings suggesting two Yb
environments.  Exploring these as well as the nature of the
intermediate ordered state are promising future steps in
single-crystal studies.

\section{Acknowledgments}
\begin{acknowledgments}
This work was supported by the US Department of Energy, Office of Basic Energy Sciences, Division of Materials Sciences and Engineering under award DE-SC0017752 (S.D.W. and M.B.). M.B. acknowledges partial support by the National Science Foundation Graduate Research Fellowship Program under grant no. 1650114. Work by L.B. and C.L. was supported by the DOE, Office of Science, Basic Energy Sciences under award no. DE-FG02-08ER46524. Identification of commercial equipment does not imply recommendation or endorsement by NIST. A portion of this research used resources at the High Flux Isotope Reactor and Spallation Neutron Source a DOE Office of Science User Facility operated by the Oak Ridge National Laboratory.
\end{acknowledgments}


\bibliography{LYO_paper1_bib_v8}

\end{document}